\documentclass[
    12pt,
    aps,
    superscriptaddress,
    nofootinbib,
    onecolumn,
    tightenlines
]{revtex4-2}

\setcitestyle{super}
\usepackage[utf8]{inputenc}

\usepackage{graphicx}
\usepackage{bm}
\graphicspath{{./figures/}}

\usepackage{textcomp,gensymb}
\usepackage{units}
\usepackage[nointegrals]{wasysym}\usepackage{dsfont}
\usepackage{placeins}
\usepackage{needspace}

\usepackage{prettyref}
\newrefformat{fig}{Fig.~\ref{#1}}
\newrefformat{exfig}{Extended Data Fig.~\ref{#1}}

\usepackage{ltablex}
\usepackage{booktabs}
\usepackage{longtable}

\newcommand{\caseboundaryrule}{%
  \specialrule{0.4pt}{1pt}{1.4pt}%
  \specialrule{0.4pt}{0pt}{1pt}%
}

\newcommand{\casefamilyfirst}[4]{%
  \multicolumn{3}{@{}p{\linewidth}@{}}{%
    \begin{minipage}[t]{\linewidth}
      \centering\textbf{#1}\par
      \vspace{1.5pt}\hrule height 0.25pt\relax\vspace{2pt}%
      \begin{tabular}[t]{@{}>{\raggedright\arraybackslash}p{0.12\linewidth}
                           >{\raggedright\arraybackslash}p{0.43\linewidth}
                           >{\raggedright\arraybackslash}p{\dimexpr0.45\linewidth-16pt\relax}@{}}
        \mbox{#2} & #3 & #4
      \end{tabular}%
    \end{minipage}%
  }\\[2pt]
}

\usepackage{xurl}
\usepackage{amsmath,amssymb}
\usepackage{mathtools}

\usepackage[normalem]{ulem}
\usepackage{color}
\usepackage{xcolor}
\usepackage{times}
\usepackage{comment}

\newcommand{\methodheading}[1]{%
  \noindent\textbf{#1}\nobreak\hspace{0.6em}\ignorespaces
}

\usepackage{hyperref}
\hypersetup{
    colorlinks=true,
    linkcolor=blue,
    filecolor=magenta,
    urlcolor=cyan,
    citecolor=red,
    pdftitle={
        A stateful research information infrastructure for scientific discovery
    },
    pdfpagemode=FullScreen,
}

\usepackage{etoolbox}
\usepackage{gewu-bibsplit}

\makeatletter

\renewcommand*\frontmatter@footnote@produce{%
  \begingroup
    \def\@thefnmark{}%
    \long\def\frontmatter@makefntext##1{\noindent##1}%
    \frontmatter@footnotetext{%
      \let\gewu@noteseparator\@empty
      \let\gewu@correspondencenotes\@empty
      \def\@TBN@opr##1##2{%
        \preto\gewu@correspondencenotes{\gewu@printcorrespondence{##1}{##2}}%
      }%
      \def\gewu@printcorrespondence##1##2{%
        \gewu@noteseparator
        \begingroup
          \csname c@\@mpfn\endcsname##1\relax
          \Hy@raisedlink{\hyper@anchorstart{frontmatter.##1}\hyper@anchorend}%
          \@textsuperscript{\frontmatter@thefootnote}##2%
        \endgroup
        \def\gewu@noteseparator{;\space}%
      }%
      \@FMN@list
      Correspondence: \gewu@correspondencenotes\par
      \noindent All authors contributed equally to this work.%
    }%
  \endgroup
  \global\let\@FMN@list\@empty
}

\makeatother

\renewcommand{\figurename}{\textbf{Fig.}}

\newcommand{\beginsupplement}{%
    \setcounter{table}{0}
    \renewcommand{\thetable}{S\arabic{table}}
    \setcounter{figure}{0}
    \renewcommand{\thefigure}{S\arabic{figure}}
    \renewcommand{\figurename}{\textbf{Supplementary Fig.}}
}

\newcommand{\Westlake}{
    Department of Physics, Westlake University, Hangzhou 310024, China
}

\newcommand{\ScienceTokyo}{
    MDX Research Center for Element Strategy, 
    Institute of Science Tokyo, Yokohama 226-8501, Japan
}

\begin{document}

\title{
    A Stateful Research Information Infrastructure for Scientific Discovery
}

\author{Lingyuan Kong}
\email{konglingyuan@westlake.edu.cn}
\affiliation{\Westlake}

\author{Zhengran Wu}
\email{wu.z.ag@m.titech.ac.jp}
\affiliation{\ScienceTokyo}

\maketitle

\begingroup

\textbf{
Artificial intelligence (AI) is being applied to an expanding range of
scientific activities\cite{wang2023scientific,messeri2024illusions,gao2024quantifying,hao2026impactfocus}, including literature analysis
\cite{dagdelen2024structured,asai2026openscholar}, hypothesis generation
\cite{ghareeb2026robin,gottweis2026coscientist,Zhang2026VirtualBiotech,sourati2023humanaware,weng2026deepscientist}, software development
\cite{aygun2026empirical,miao2026paper2agent,lu2026aiscientist} and automated experimentation
\cite{shi2026qumus,qiu2026aims,Bauer2026AutonomousQuantumTransport,Cheng2026UnmodeledStates,boiko2023coscientist,angello2022closedloop}.
Yet these capabilities rely on a more fundamental condition: scientific
information must remain identifiable and traceable 
across systems and over time
\cite{moreau2015prov,wilkinson2016fair,rauber2015dynamic}.
In practice, different versions of a work and their associated records
and files are often scattered across services and local directories
\cite{niso2008jav,boardman2004crosstool}.
This fragmentation forces researchers to identify and contextualize the same
work repeatedly, disrupting continuity from discovery to interpretation and
reuse\cite{whittaker2011pim}.
Here we present Gewu Insight\cite{Wu2026gewuinsight}, a research
information infrastructure that addresses this discontinuity by organizing
scientific works as persistent, stateful research objects
\cite{belhajjame2015researchobjects,soilandreyes2022rocrate}.
It unifies venue-, topic- and author-based discovery records through
a shared paper identity. Within a local library, metadata,
versioned resources and research context can then accumulate around
the same work.
These capabilities are integrated into an iterative workflow for scientist-led, AI-assisted research, allowing earlier interpretations to be revisited and refined.
A bounded evaluation showed high candidate precision across the three discovery modes, with paper identity, resource distinctions and research state preserved in controlled workflows.
By making continuity an explicit property of research infrastructure,
Gewu Insight 
can serve as an inspectable scientific memory layer shared by researchers and AI assistants, supporting human judgment and providing a foundation for future, more epistemically grounded scientific AI
\cite{krenn2022understanding,messeri2024illusions,bottino2026retrieval}.
}

\vspace{5pt}

Scientific information does more than support individual research tasks; it
connects them\cite{yang2025airalogy}. A paper found during discovery may later inform a hypothesis, a
computational procedure or an experimental decision\cite{pagel2026chemputation,pyzerknapp2022accelerating}
(Fig.~\ref{fig:fig1}a). 
Ambiguity in its identity, provenance, version or publication status can
therefore propagate through the research process
\cite{hsiao2021continued,thelwall2025ignore,sambasivan2021cascades}, making information continuity a
prerequisite for traceable scientific reasoning rather than merely a matter
of document management
\cite{moreau2015prov,wilkinson2016fair,wang2026physmatai}.
Maintaining this continuity requires distinguishing a scientific work
from its individual records and files. A preprint and its published version
report the same research but may differ in content, while supplements and
other attachments serve different purposes. Rediscovery or reuse in another
project should add context to the existing work rather than create another
paper object. A research system must therefore keep these versions,
resources and contexts distinct while linking them to one work identity
\cite{niso2008jav,riva2017iflalrm,moreau2015prov}.

We use the term \emph{stateful research information infrastructure} to
describe a system that makes this continuity explicit. Here,
\emph{stateful} means that a scientific work persists as an identifiable
object across sources, interfaces and time, while its metadata, provenance,
resources, research contexts 
and durable operational
state remain associated with it
\cite{belhajjame2015researchobjects,moreau2015prov,wilkinson2016fair}.
The term therefore refers to object-level continuity rather than merely to the persistence of a software session.
The system specifies what each operation changes and retains. This keeps sources, retained versions and analyses connected and accessible through rediscovery, version changes and reuse across projects, providing a continuing basis for human judgment and AI-assisted inquiry.

\setlength{\textfloatsep}{8pt plus 1pt minus 1pt}

\begin{figure}[t]
\centering
\includegraphics[width=\linewidth]{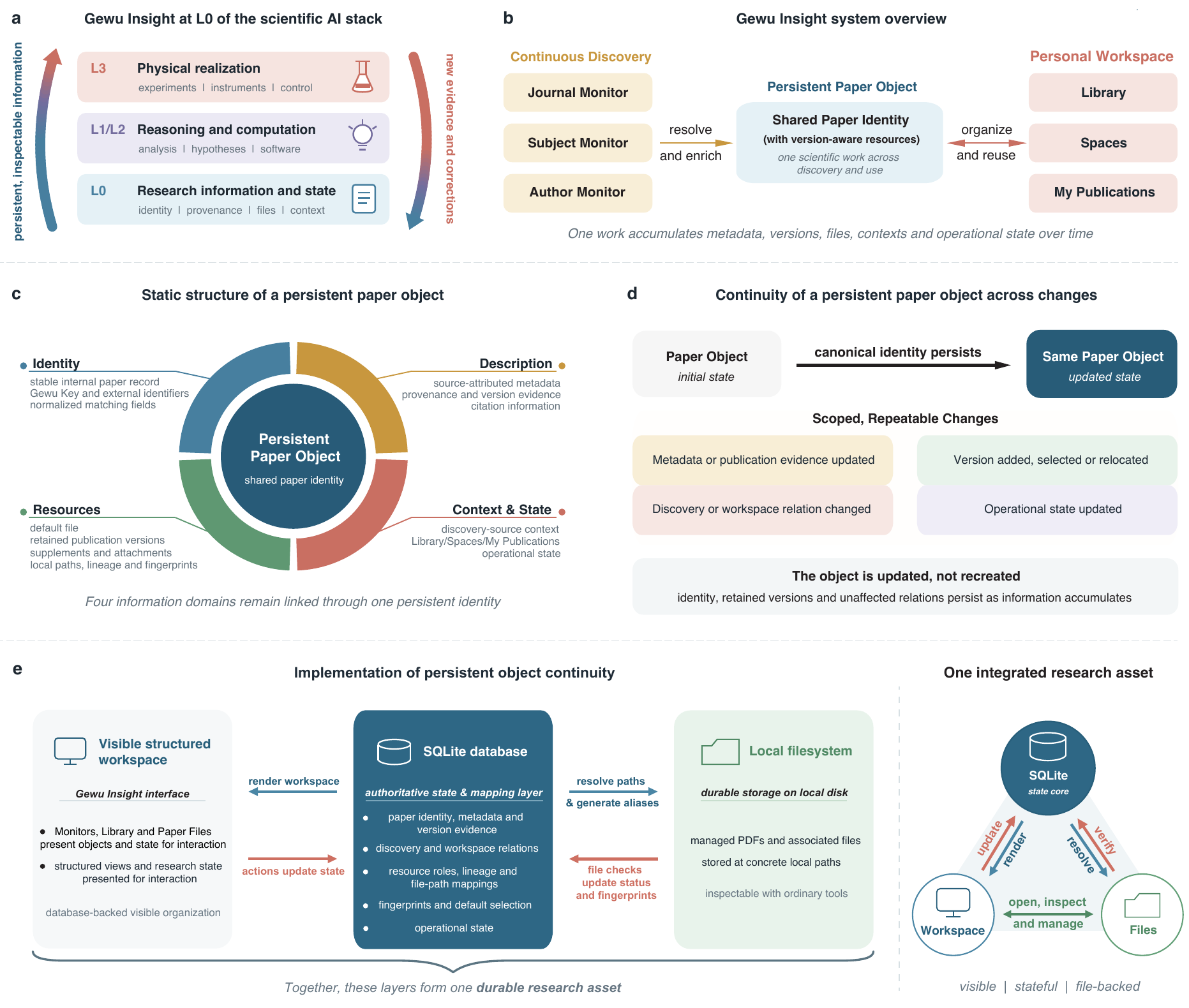}

\begingroup
\setlength{\abovecaptionskip}{0pt}
\setlength{\belowcaptionskip}{0pt}
\small
\caption{
\textbf{A persistent-object architecture for scientific information continuity.}
\textbf{a}, A layered view of scientific AI: research information and state
(L$_0$), reasoning and computation (L$_1$/L$_2$), and physical realization
(L$_3$). Gewu Insight operates at L$_0$, providing persistent,
inspectable information to higher layers and incorporating new evidence
and corrections.
\textbf{b}, A shared paper identity and version-aware resources connect
continuous discovery through Journal, Subject and Author Monitor with
organization and reuse in Library, Spaces and My Publications, allowing
information and research state to accumulate around one work.
\textbf{c}, Four distinct domains share one persistent identity: paper
identity, source-attributed description, resources, and context and state.
Resource records describe publication versions and associated files,
including available local paths, lineage and fingerprints.
\textbf{d}, Changes across these domains update the paper object without
recreating it, preserving its canonical identity, retained versions and
relations outside each operation's scope.
\textbf{e}, The workspace displays authoritative state held in SQLite,
while user actions update it. Stored paths link database records to managed
files in the local filesystem. Together, workspace, database and files form
an inspectable, reusable and durable research asset.
}
\label{fig:fig1}
\endgroup
\end{figure}

\vspace{2pt}

\begingroup
\raggedbottom
\setlength{\skip\footins}{13pt plus 2pt minus 2pt}
\renewcommand{\footnoterule}{\kern-6pt\hrule width .5in\kern6pt}
\Needspace*{4\baselineskip}
\endgroup
\noindent {\bf Persistent Object Architecture}

Discovery and personal organization converge on a
persistent paper object (Fig.~\ref{fig:fig1}b) in Gewu Insight. Each object is assigned a Gewu Key, a persistent identifier that remains
stable as metadata, publication files and research contexts accumulate.
The object is organized into four connected information domains
(\prettyref{fig:fig1}c). 
First, identity information includes the internal paper record, available
bibliographic identifiers and normalized fields for matching records
across sources. 
Second, descriptive evidence
preserves source-attributed metadata, provenance, publication evidence and
citation information. 
Third, resource relations, exposed through Paper Files in the Library, connect
the paper to publication versions, supplements and other attachments,
recording the Default selection, provenance and local paths.
Fourth, context and state record the paper's
relations to discovery sources (Supplementary
Fig.~\ref{fig:discovery-interface}), Library folders, Spaces and My Publications (Supplementary
Fig.~\ref{fig:interface}),
together with relevant operational state (capitalized component names denote elements of the
Gewu Insight interface; see Methods).

Continuity depends on specifying what each operation changes and what it must
retain (\prettyref{fig:fig1}d). A metadata correction updates descriptive
evidence without changing paper identity; selecting a publication file changes
the Default without removing retained versions; and adding or removing a
project association leaves other uses of the paper intact. These distinctions
allow subsequent reading and analysis to return to retained materials even as
the paper's description, files and organization change.

At the implementation level, these principles are realized through the
interaction of the application interface, an embedded SQLite relational
database\cite{sqlite}, and the local filesystem
(\prettyref{fig:fig1}e; see Methods and Supplementary
Fig.~\ref{fig:storage-mapping}). SQLite is the system of record for paper identities,
metadata and publication evidence, discovery and workspace relations,
resource and operational state, and the paths linking records to local files. 
The interface displays this structured state, and user actions update
the corresponding database records. Portable Document Format (PDF) files, supplementary materials and imported
attachments remain ordinary filesystem objects that can also be inspected
with external tools. Stored paths and filesystem aliases connect the
structured workspace to these underlying documents. This arrangement preserves paper identity and its associated sources,
resources and research contexts across navigation, file operations and
application restarts.

\begingroup
\setlength{\intextsep}{16pt plus 1pt minus 1pt}

\begin{figure}[h]
\centering
\includegraphics[width=\linewidth]{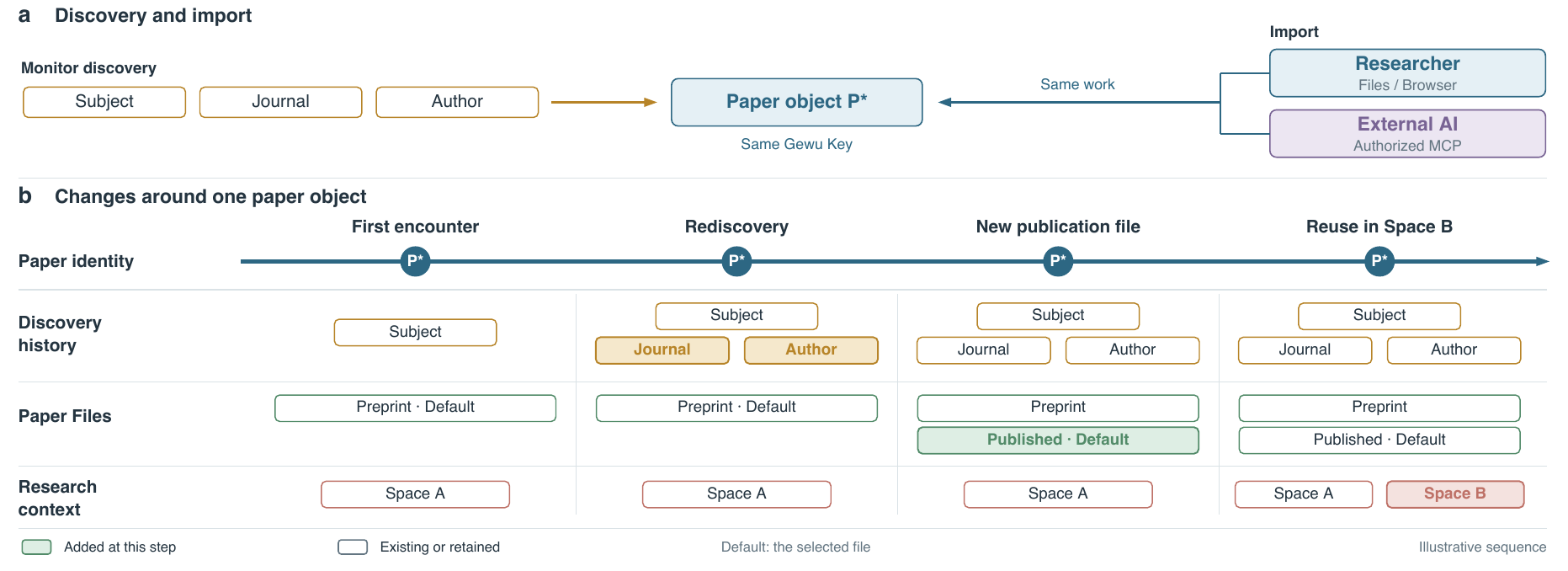}

\setlength{\abovecaptionskip}{10pt}
\setlength{\belowcaptionskip}{0pt}
\small
\caption{
\textbf{Discovery and import converge on one persistent paper object.}
\textbf{a}, Journal, Subject and Author Monitor provide complementary discovery
routes. Researcher-directed imports and authorized Model Context Protocol
(MCP) imports by external AI clients also extend the same paper object when the work is already known,
reusing its Gewu Key.
\textbf{b}, An illustrative sequence follows the same paper object
(P$^\ast$) through rediscovery, publication-file addition and reuse in another
Space. New discovery records accumulate, a published file is selected as
Default while the preprint remains, and Space B is added without removing
Space A. Filled labels mark additions at each step; outlined labels mark
existing or retained items.
}
\label{fig:fig2}
\end{figure}

\endgroup

\noindent {\bf From Discovery to Reuse}\par\nopagebreak[4]

Researchers follow publication venues, research topics and individual
scientists to keep up with an evolving literature. Gewu Insight supports
these complementary practices through Journal, Subject and Author Monitor
(\prettyref{fig:fig2}a; see Methods and Supplementary
Fig.~\ref{fig:discovery-interface}). Researchers and external AI clients can
also import acquired material through the application or authorized Model
Context Protocol (MCP) connections, respectively. Imports identified as the same work reuse its
existing Gewu Key (see Methods). A paper first encountered through a subject
search may later appear in a journal or among a monitored scientist's
publications. Identity resolution links these encounters to the same paper
while retaining the source and observation context of each
route\cite{moreau2015prov}. New routes therefore extend the paper's discovery
history even when the work is already known.

The record can then accumulate metadata, publication versions and associated
files without replacing the paper object. In the illustrated sequence, a
published file is added and selected as Default while the preprint remains;
reuse in Space B adds an association without removing Space A
(\prettyref{fig:fig2}b). Library folders, Spaces and My Publications provide
organizational contexts for such use (Supplementary Fig.~\ref{fig:interface}).
When the researcher returns to the paper, its discovery history, earlier
project contexts and retained files remain available
\cite{whittaker2011pim,belhajjame2015researchobjects}.

\vspace{10pt}
\begingroup
\setlength{\intextsep}{8pt plus 1pt minus 1pt}

\begin{figure}[h]
\centering
\includegraphics[width=\linewidth]{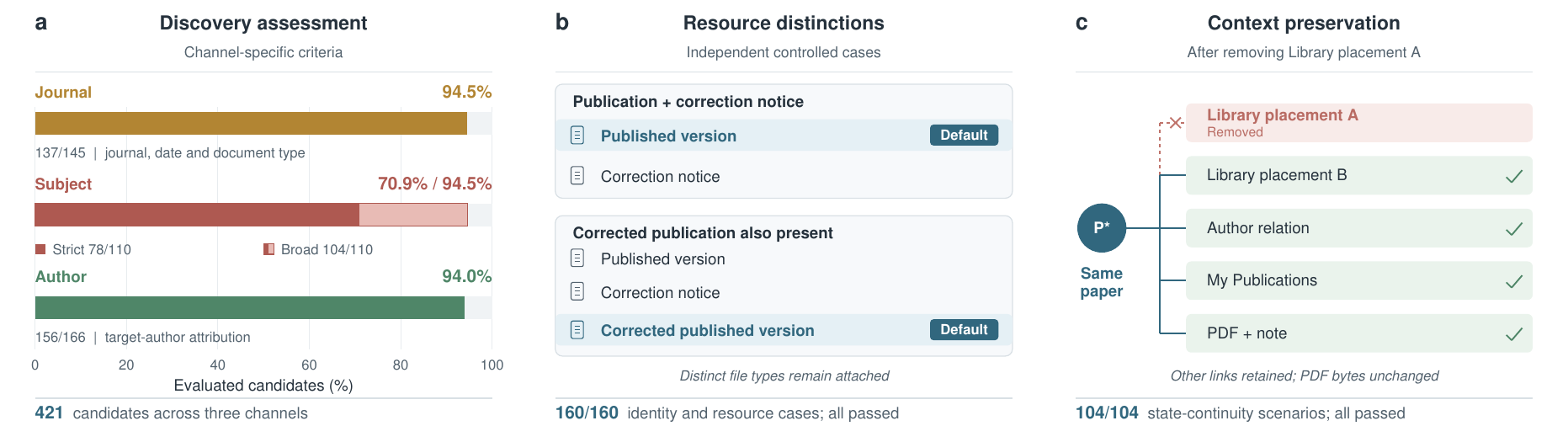}

\setlength{\abovecaptionskip}{2pt}
\setlength{\belowcaptionskip}{2pt}
\small
\caption{
\textbf{Evaluation of discovery and information continuity.}
\textbf{a}, Discovery assessment by channel: joint journal, date and document-type
eligibility, strict and broad topical relevance, and target-author attribution.
Fractions report confirmed candidates. The Subject bar nests strict relevance
(dark) within broad relevance (dark plus light), using the same 110 candidates.
{After screening for document scope and availability of verifiable
source evidence, 166 of 233 sampled Author Monitor candidates were evaluated,
with target-author attribution confirmed for 156/166 (94.0\%; Supplementary
Table~\ref{tab:discovery-screening}).}
\textbf{b}, Two independent controlled cases with specified file types
illustrate resource distinctions and automatic Default selection. A correction
notice remains separate from the published version. When a corrected published
version is also present, it becomes Default while the other resources remain
attached. All 160 predefined identity and resource cases passed.
\textbf{c}, Removing one Library placement preserves the paper's identity,
other associations and both resources, with the published PDF's bytes
unchanged. P$^\ast$ denotes the same paper object; the crossed dashed link
marks the removed placement, and check marks indicate retained relations or
resources. All 104 complete state-continuity scenarios passed.
Panels \textbf{b} and \textbf{c} show selected examples from the evaluation
(see Methods and Supplementary Fig.~\ref{fig:evaluation-coverage}b,c).
}
\label{fig:fig3}
\end{figure}

\endgroup

\noindent {\bf Discovery Quality and Information Continuity}

We evaluated discovery quality, identity and resource handling, and
information continuity during subsequent use
(\prettyref{fig:fig3}; Supplementary Fig.~\ref{fig:evaluation-coverage}a; see Methods).
Journal candidates met the journal, date and document-type criteria in
137/145 cases (94.5\%), and target-author attribution was confirmed in
156/166 cases (94.0\%). Subject candidates showed strict topical relevance
in 78/110 cases (70.9\%) and broad relevance in 104/110 (94.5\%) in a supplementary assessment, including
papers with substantive links to the topic even when it was not their main focus.
These proportions describe discovery quality within the evaluated sample
and channel-specific criteria.

In controlled tests, the system resolved paper identity while keeping
the different resources associated with a work distinct. All 160 controlled identity and
resource cases passed, including cases in which conflicting evidence
prevented automatic merging
(Supplementary Fig.~\ref{fig:evaluation-coverage}b).
For example, a correction notice remained attached to the paper but did not
replace the published article as the Default file. In a separate case, a
corrected published version became Default while the other files remained
available (\prettyref{fig:fig3}b).

Information continuity was also preserved as papers were rediscovered,
updated and reorganized. All 104 controlled workflow scenarios passed
(Supplementary Fig.~\ref{fig:evaluation-coverage}c).
Rediscovery added another discovery route to the existing paper rather
than creating a duplicate, and adding a published version preserved the
earlier preprint. Removing one Library placement left the paper's other
associations and both files intact (\prettyref{fig:fig3}c). Within these
tested workflows, changes to one part of a paper's record therefore left
the information associated with its other uses unaffected.

\vspace{5pt}
\noindent {\bf Facilitating AI-assisted Research}

Gewu Insight supports scientist-led, AI-assisted research by anchoring
successive analyses to persistent paper objects. Through the same Gewu Key,
researchers and authorized AI assistants can return to a paper's retained
sources and prior analyses as questions, research groupings and external
syntheses change. We illustrate this workflow through a literature study of
intermediate-band solar cells (IBSCs), asking how experimental observations
support a two-step absorption mechanism and how that evidence relates to
device performance (\prettyref{fig:fig4}). Guided by the researcher, external
AI assistants brought together existing holdings and materials obtained
through searches and citation tracing in a dedicated Space (see Methods).

\vspace{2pt}
\begingroup
\setlength{\intextsep}{8pt plus 1pt minus 1pt}

\begin{figure}[h]
\centering
\includegraphics[width=\linewidth]{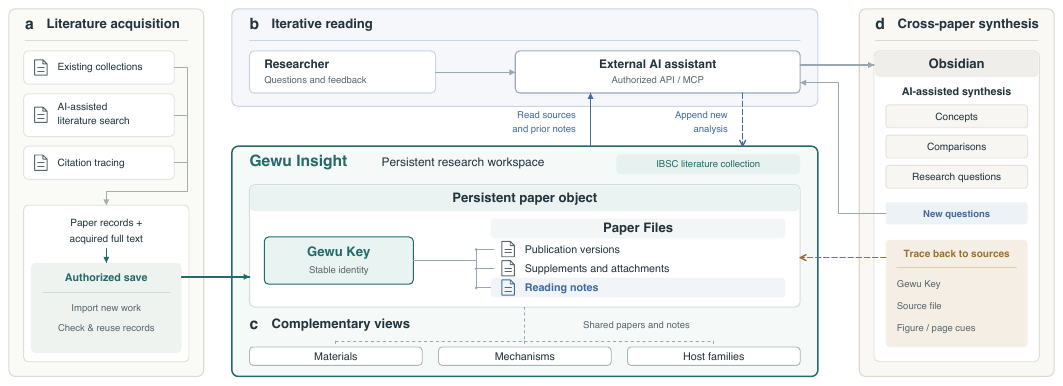}

\setlength{\abovecaptionskip}{2pt}
\setlength{\belowcaptionskip}{2pt}
\small
\caption{
\textbf{A persistent workspace for AI-assisted research.}
Gewu Insight connects literature acquisition, iterative reading,
reorganization and cross-tool synthesis around persistent paper objects.
\textbf{a}, Existing collections, AI-assisted searches and citation tracing
supply paper records and acquired full text for authorized saving and record
reuse.
\textbf{b}, Researcher questions and feedback guide iterative reading: external
AI assistants revisit sources and prior notes linked by the same Gewu Key and
append new analysis to that paper.
\textbf{c}, Materials, mechanisms and host families provide complementary views
of shared papers and reading notes.
\textbf{d}, AI-assisted synthesis in Obsidian\cite{obsidian} connects concepts, comparisons and
research questions to evidence held in Gewu Insight. New questions can return to the
assistant for further source examination; retained Keys, source-file
references and figure/page cues preserve routes back to evidence. The intermediate-band solar cell (IBSC)
collection illustrates this workflow; the connections represent authorized
research actions rather than native automatic synchronization.
}
\label{fig:fig4}
\end{figure}

\endgroup

To examine whether sub-bandgap
photocurrent was sufficient to establish two-step absorption, the AI assistant
retrieved Cheriton et al.'s paper\cite{Cheriton2020TwoPhoton} and its existing
analysis, then revisited Fig.~3 and Supplementary Fig.~6 of that study. Rereading clarified
that the response without light bias was insufficient by itself, that the
light-bias-induced increment needed to be assessed alongside temperature and
weak-light controls, and that part of the room-temperature enhancement
remained unexplained. Gewu Insight appended these clarifications to the same paper
record while retaining the source and earlier analysis, making the refined
understanding and its evidential limits available to subsequent research
(\prettyref{fig:fig4}b; Supplementary Table~\ref{tab:ibsc-interpretation-trace}).

When research turns to cross-paper comparison, different groupings in Gewu Insight
can draw on the same papers and accumulated analyses, allowing reading
undertaken for one question to inform others. IBSC papers grouped by formation
mechanism and host family provided a basis for the researcher and AI to compare
Cheriton et al.'s study with that of Sogabe et al.\cite{Cheriton2020TwoPhoton,Sogabe2023CurrentConstraint}.
The former provided experimental support for two-step photocurrent, while
overall device efficiency remained very low; the latter achieved a higher
efficiency under concentrated illumination but had not independently
determined the contribution of two-step absorption to output power. The
comparison clarified an evidential gap between mechanism verification and
performance attribution: observing a two-step process does not directly
establish a net power gain, and a higher device efficiency does not by itself
identify the source of that gain. Gewu Insight retained the associations among each
paper, its analyses and research groupings, allowing the experimental
conditions and evidential limits behind these judgments to be revisited
across comparison tasks (\prettyref{fig:fig4}c).

Researchers can also organize accumulated understanding in external knowledge
workspaces while retaining routes back to sources and analyses in Gewu Insight. In
this case, Gewu Insight held the source papers and their individual analyses, while
the researcher and AI \mbox{organized} cross-paper concepts, comparisons and research
questions in a knowledge-management tool (Obsidian\cite{obsidian}),
linking them to evidence through
citations, Gewu Keys and figure or page locations (\prettyref{fig:fig4}d).
The comparison led to a further question: which measurements are needed to
connect evidence for two-step absorption to its contribution to output power?
The synthesis note retained this question together with its basis, allowing
subsequent tasks to follow the references back to Gewu Insight, retrieve the sources
and prior analyses, and continue the investigation.

\FloatBarrier

\vspace{5pt}
\noindent {\bf Discussion}

Reference managers already support reuse of a single item across
collections\cite{zoteroCollections}, while research-object and provenance
approaches describe relationships among research materials and their
origins\cite{soilandreyes2022rocrate,huber2020aiida}.
Gewu Insight brings these principles into a continuous literature workflow,
connecting discovery, version updates, project reuse and successive human-AI
analyses through a shared paper identity and explicit rules for what each
operation changes and retains. Together, these rules address a
recurring problem: rediscovering a paper does not by itself reconnect the
researcher or AI assistant to the versions, research contexts and analyses
retained during earlier work. Both can return to this connected record as
questions and tools change. Controlled workflows verify preservation of
identity, resources and unaffected associations through these changes
(\prettyref{fig:fig3}), while the IBSC study illustrates how this shared record
supports renewed source examination and cross-paper comparison
(\prettyref{fig:fig4}).

The stateful workspace also provides a shared basis for human-AI
collaboration. Large volumes of AI-generated information can be difficult to
organize and revisit\cite{shao2026sciscigpt,suh2023sensecape}, while evaluating synthesized interpretations requires
access to their sources\cite{kang2023synergi}.
Gewu Insight organizes papers, retained versions and analyses for
researchers to inspect through a visible workspace and authorized AI
assistants to access through structured interfaces. 
The workspace thus supports
scientist-led, AI-assisted research, with researchers directing questions and
assessing interpretations\cite{messeri2024illusions}.

The same scientist-led approach shaped the development of
Gewu Insight. The project began with literature monitoring and evolved
through repeated use into a workspace for retaining and reusing research
materials. Researchers specified the required behavior and assessed working
versions, while AI coding assistants helped implement and revise the design.
Practical expectations about how papers and their associated information
should be retained became explicit rules in the data model, with automated tests checking that software updates continued to meet these requirements
(see Methods and Supplementary Fig.~\ref{fig:development}). This experience
suggests that AI-assisted development can help researchers without formal
software-engineering training build and adapt tools around
their own practice, with domain knowledge guiding both what is built and how
it is assessed.

A natural extension is to connect interpretations more precisely to their
supporting evidence. Persistent paper identities provide a starting point
for linking individual claims to the versions, figures or passages that
support them. Nanopublications and scholarly knowledge graphs offer
approaches to representing such relations beyond the document level
\cite{groth2010nanopublication,jaradeh2019orkg,ran2026asks}. When sources change, such
links could help identify which interpretations warrant reconsideration.
Retaining those revisions alongside their supporting evidence would make
the development of an interpretation inspectable, rather than recording only
its latest form. Together, persistent scientific memory and scientist-led
tool development point towards more epistemically grounded science, in which
researchers can continue to inspect evidence, revise interpretations and
govern their tools.

\endgroup

\raggedbottom

\let\oldaddcontentsline\addcontentsline
\renewcommand{\addcontentsline}[3]{}

\makeatletter
\patchcmd{\bibsection}{\onecolumngrid@push}{}{}
  {\PackageError{gewu}{Unable to remove bibliography grid push}{}}
\patchcmd{\bibsection}{\onecolumngrid@pop}{}{}
  {\PackageError{gewu}{Unable to remove bibliography grid pop}{}}
\makeatother
\def\urlprefix{}
\bibliographystyle{gewu-naturemag}
\let\GewuSavedBibFont\bibfont
\apptocmd{\bibfont}{\setlength{\baselineskip}{14pt}}{}
  {\PackageError{gewu}{Unable to adjust main reference line spacing}{}}
\addtolength{\bibsep}{-1.5pt}
\bibliography{Gewu_BibTeX}
\addtolength{\bibsep}{1.5pt}
\let\bibfont\GewuSavedBibFont

\let\addcontentsline\oldaddcontentsline

\clearpage
\flushbottom

\beginsupplement

\renewcommand{\theHfigure}{supp.\arabic{figure}}
\renewcommand{\theHtable}{supp.\arabic{table}}

\begingroup

\noindent {\bf \large Methods}

\vspace{6pt}

\methodheading{Research interface and workspace organization}
Library provides an organized collection for long-term scientific memory
and retrieval, while remaining open to new additions
(Supplementary Fig.~\ref{fig:interface}a).
Spaces are independent workspaces at the same organizational level as
Library, rather than folders within it. They are intended for active
topics or projects, where new material is added as work progresses
(Supplementary Fig.~\ref{fig:interface}b). Once a project is complete,
the user may move its Space into Library.
Within Library and Spaces, Paper Files exposes the publication versions
and associated resources of the selected paper. My Publications collects
the user's own papers and displays authorship roles and source-specific
citation information. Its Export List function produces a detailed,
well-structured publication list customized to the user's needs
(Supplementary Fig.~\ref{fig:interface}c).
The three Monitor interfaces support complementary discovery practices
(Supplementary Fig.~\ref{fig:discovery-interface}).
Journal Monitor follows publication venues and provides calendar and
journal filters. Subject Monitor presents candidates for configured
research topics together with their matched keywords. Author Monitor
follows selected researchers' publication lists and identifies subsequent
additions. Candidates can then be added to Library for organization and reuse.

\vspace{2pt}
\methodheading{System and data architecture}
Gewu Insight combines a desktop application, a local Python
service, an embedded SQLite relational database \cite{sqlite}, and
a managed but user-visible filesystem (Fig.~\ref{fig:fig1}e).
Primary Library and state-management functions operate locally without
a remote application server. SQLite is the authoritative store for
structured state, with write-ahead logging and versioned schema
migrations. The Browser workspace supports webpage and PDF browsing;
adding a paper from the Browser requires an explicit save action.

Each canonical paper is assigned a {Gewu Key}, an application-specific
identifier that provides a stable reference to the paper object rather
than to an individual file (Supplementary Fig.~\ref{fig:interface}a).
External bibliographic identifiers, such as digital object identifiers (DOIs) and arXiv identifiers, are
recorded separately when available. When duplicate paper records are
reconciled, earlier Gewu Keys remain resolvable through aliases.
During identifier-based resolution, conflicting bibliographic evidence
is recorded for review rather than accepted as an automatic merge.

The data model separates source observations, organizational relations
and file resources\cite{riva2017iflalrm,belhajjame2015researchobjects,moreau2015prov} (Supplementary Fig.~\ref{fig:storage-mapping}a;
Supplementary Table~\ref{tab:sqlite-schema} lists the database tables
and their roles). Source observations retain the source and first- and
last-observed times; repeated encounters update the same source relation.
Separate relations connect the paper to subjects, researchers and workspace
locations. Resource records retain file type, source, annotation state,
local paths, fingerprints and available lineage evidence. Default selection
is maintained at the paper level and determines which associated file is
opened by default.

The filesystem mapping preserves a visible organization of these resources
(Supplementary Fig.~\ref{fig:storage-mapping}b). Library folders and
Spaces map to corresponding directory hierarchies, while My Publications
has a dedicated directory. An {Associated Files} tree mirrors the
workspace hierarchy, with paper-specific folders bearing the Gewu Key
and subfolders for versions, supplements, figures and other files.
Additional workspace placements can link to the same primary PDF.
Paper Files brings resources together through their database relations
and recorded paths, rather than requiring every file to share one
directory. Changing the Default PDF can relocate the selected file and
retain the previous primary file under {Versions}, updating paths
and links without changing file contents.

Folder Loader copies imported files into managed storage while retaining
their source paths and relative organization, leaving external originals
unchanged. Removing a discovery relation, a workspace placement or a
physical resource has a distinct scope. SQLite transactions protect
database updates; file-reassignment operations additionally track file
moves and alias changes for rollback if the operation fails.

\methodheading{Discovery and metadata acquisition}
The three Monitors provide complementary acquisition routes
(Supplementary Fig.~\ref{fig:discovery-interface}).
Journal Monitor retrieves papers from configured journal sources and
user-selected arXiv\cite{arxivapi} archive groups. Subscriptions and discovery scopes
are stored in SQLite, and results are organized by publication or
announcement date.

Subject Monitor combines locally stored arXiv\cite{arxivapi} records and cached journal
feeds with bounded queries to OpenAlex\cite{priem2022openalex}, Crossref\cite{hendricks2020crossref}, Semantic Scholar\cite{kinney2023semanticscholar} and
PubMed\cite{pubmedabout}. Source records are normalized before matching against user-defined
subjects and keyword groups. Filters exclude non-target article types,
including correction and retraction notices, editorials and news, as well
as datasets, software and other non-literature records. Zenodo\cite{zenodo2013} records
are excluded during both ingestion and cached-result display. These
filters govern candidate eligibility, not comprehensive tracking of
editorial-status changes to previously stored papers.

Author Monitor establishes an initial publication baseline for a selected
researcher and identifies subsequent additions. Google Scholar\cite{googlescholarprofiles} profiles
are queried directly. For ResearchGate\cite{researchgateprofile} profiles, publication records are
matched through OpenAlex\cite{priem2022openalex}, with Semantic Scholar\cite{kinney2023semanticscholar} as a fallback. New
observations add researcher-paper relations rather than independent
copies of an existing work.

Metadata enrichment draws on Crossref\cite{hendricks2020crossref}, arXiv\cite{arxivapi}, PubMed\cite{pubmedabout}, the Astrophysics Data System (ADS)\cite{kurtz2000ads}, OpenAlex\cite{priem2022openalex} and
Semantic Scholar\cite{kinney2023semanticscholar} for available identifiers, publication information and
citation metadata.
DataCite\cite{dataciteapi} provides an additional DOI lookup route where repository records
are permitted. Publisher pages supply further publication and file
evidence.

Discovery records retain their source and observation context
\cite{moreau2015prov}. Subject Monitor additionally records per-source
runs and query outcomes, distinguishing completed searches, including
empty results, from partial, deferred or failed retrievals. Journal and
Author Monitor retain their corresponding fetch or profile state.
Partial acquisitions preserve already committed papers and their
provenance. Metadata refreshes enrich existing records without removing
workspace relations, personal-publication fields or local resources.

The Browser provides a user-directed acquisition route. Ordinary
navigation, bookmarks and browsing history do not create paper records.
An explicit save action presents captured metadata, candidate
files and a Library or Space destination for review. Confirmed captures
enter the same paper and resource workflows as other acquisition routes,
with PDFs validated before registration and Browser provenance retained
as a source observation.

\vspace{2pt}
\methodheading{Identity, version and resource resolution}
DOI representations are normalized, and arXiv version suffixes are
separated from the base identifier.
Identifier matches are checked for DOI/URL inconsistencies
and combined conflicts in title, authorship and journal evidence.
Conflicting observations are recorded for review without automatic merging;
in this identifier-based path,
title agreement alone is insufficient to establish identity (Supplementary
Table~\ref{tab:task-b-cases}, B1).
Author Monitor also permits a heuristic exact-title/year fallback
without independent authorship confirmation.
Missing identifiers may be recovered from a local
PDF to guide metadata lookup, while sparse source responses do not
erase existing metadata or personal fields.

Paper Files records file type, acquisition source and annotation state
independently (Supplementary Fig.~\ref{fig:interface}a), distinguishing
publication versions from supporting materials without conflating
publication status with provenance or reader modification
\cite{niso2008jav,crossrefversioning}. Classification follows evidence from the actual
file and its source, rather than the paper's latest bibliographic
metadata. Automatic Default selection prioritizes corrected published
versions, published versions, accepted manuscripts and preprints,
in that order. Supporting materials and invalid files are excluded
from this ranking. A first Browser-captured article PDF of unknown
type can serve as a fallback when no Default exists. A manually
selected file overrides automatic selection while it remains available
and is not classified as invalid, until the user restores Automatic mode.

Exact file hashes, computed with Secure Hash Algorithm 256 (SHA-256), identify byte-identical files for reuse. An
annotation-insensitive PDF content fingerprint can relate clean and
annotated copies while preserving them as separate resources. It hashes
decoded page-content streams, page geometry and rotation, and image and
form objects, excluding PDF annotation entries. Automatic
annotation detection examines reader-authored markup rather than
file-size changes. The content fingerprint is advisory: it does not
alone justify merging or deleting files, or establish their scientific
equivalence.

PDF downloads are staged in temporary files and registered only after
structural validation, including header and end-marker checks, parser
readability and a non-zero page count. Password-locked or malformed
inputs are rejected without replacing an existing file or registering
a partial resource. Valid new versions remain associated with the same
paper alongside previously retained files.

\vspace{2pt}
\methodheading{AI access and analysis persistence}
Authorized AI clients work with the same paper records and resources as
researchers through an opt-in local application programming interface (API) and a Model Context Protocol (MCP)
adapter\cite{mcp2025specification} (Fig.~\ref{fig:fig4}). Each connection specifies its collection and
tool permissions (Supplementary Table~\ref{tab:mcp-capabilities}). The local
service checks permissions for file reading, analysis saving, import and
organization. Clients identify papers by Gewu Key and can
obtain read-only copies of registered files through MCP for inspection.
Model inference is handled by the external client rather than by Gewu Insight itself.

Analyses saved through this interface are appended as Markdown resources in
Paper Files, leaving source files, earlier analyses and any existing Default
selection unchanged.
Each note records its Gewu Key and originating connection.
Source-file linkage is optional; when specified, it requires both the file
identifier and its SHA-256 hash, while page references remain optional.
Before accepting a new file-linked analysis, the service recomputes the
source hash and rejects a mismatch with a request to read the file again.
Notes saved without a source-file link are not subject to this check. SQLite records the save
receipt and resource relation, while the analysis text remains in the local
Markdown file (Supplementary Table~\ref{tab:sqlite-schema}). Cross-tool
updates require explicit client actions rather than automatic synchronization.

\vspace{2pt}
\begingroup
\widowpenalty=10000
\methodheading{IBSC literature-study workflow}
The IBSC case examined how sub-bandgap photoresponse supports two-step
absorption and whether that evidence establishes the process's contribution
to output power. Researchers and external AI assistants assembled existing
holdings and materials obtained through searches and citation tracing in
the IBSC Space in Gewu Insight, retaining source papers and individual reading notes.
Groupings by material, formation mechanism and host family provided
complementary reading routes. For this question, Cheriton et al.'s
InGaN/GaN nanowire study and Sogabe et al.'s InAs/GaAs quantum-dot device study
were selected for closer examination of their experiments and measurement
conditions alongside existing analyses.
The Cheriton rereading and comparison with Sogabe et al. were carried out
using Codex Desktop (recorded CLI component version
0.155.0-alpha.16.3) with model identifier GPT-6-Astra.

Single-paper rereading asked whether sub-bandgap response alone established
two-step absorption. The existing Cheriton analysis already distinguished
photocurrent enhancement from power conversion efficiency. Follow-up reading
examined the controls in the paper's Fig.~3 and Supplementary
Fig.~6\cite{Cheriton2020TwoPhoton}. Sub-bandgap response was present without
850-nm light bias; the authors attributed this baseline to single-photon
leakage, potentially through defect or surface states. The baseline therefore
did not uniquely identify a two-step process. The light-bias-induced
increment, considered with temperature and weak-light controls, supported
two-step photocurrent. Rereading clarified that Fig.~3e measured temperature
dependence without light bias, rather than directly measuring laser-induced
heating. Weak-light measurements further argued against substantial heating
as the dominant explanation, but did not explain the complete room-temperature
response, particularly the enhancement above the host bandgap. The new
analysis retained these limits and recorded the reported maximum 44\%
relative photocurrent enhancement separately from the maximum 0.016\%
power conversion efficiency measured independently from current-voltage
characteristics at 4 suns.

Cross-paper comparison then examined whether a high device efficiency
established the power contribution of two-step absorption. Sogabe et al.
combined an InGaP top cell with an InAs/GaAs quantum-dot bottom cell in a
series-connected structure and reported 22.8\% power conversion efficiency
at 93 suns\cite{Sogabe2023CurrentConstraint}. Rereading focused on quantum-dot
spectral response, efficiency under concentration and device controls.
Comparisons in the paper's Fig.~6c involved different antireflection-coating
conditions, so the efficiency differences could not be attributed entirely
to the intermediate band. The authors also related fill-factor recovery under
concentration to subcell current limitation and interface-related shunting.
These results supported the reported whole-device performance but did not
independently isolate the power contribution of two-step absorption.
The studies thus provided complementary evidence: Cheriton et al. offered
more targeted experimental support for two-step photocurrent while device
efficiency remained low; Sogabe et al. achieved higher concentrated-light
efficiency, but that value alone did not establish the mechanism responsible.

New analyses were saved in the corresponding papers' Paper Files, recording
the experiments examined, measurement conditions, figure or page locations
and interpretive limits. Cross-paper comparisons and research questions were
recorded in an Obsidian synthesis note, linked to sources and analyses in
Gewu Insight through citations, Gewu Keys, source-file identifiers and figure or page
locations.
Supplementary Table~\ref{tab:ibsc-interpretation-trace} traces
one retained Cheriton record from the earlier analysis through source rereading
and an appended follow-up to its explicit citation in the Obsidian synthesis.

The synthesis retained questions about connecting selective two-beam response
to maximum-power changes in the same device at matched temperatures and
voltage operating points; controlling antireflection coatings and interface
processing; and distinguishing changes in heating, recombination and carrier
collection. Additional auxiliary illumination also requires accounting for
its input energy. These questions and their literature basis were retained
in the synthesis note.
\par
\endgroup

\vspace{2pt}
\methodheading{Scientist-led development}
Development proceeded through iterations of research use, implementation and
checking. Researchers described the required behavior and assessed working
versions; AI coding assistants helped translate those requirements into
changes to the data model and interface. Feedback from use and testing
informed subsequent revisions. For example, the requirement to remove a
paper from one project without disrupting access elsewhere led to a
distinction between workspace placement and file ownership, with regression
checks for retained associations and file access.

We reconstructed this process from dated design discussions, versioned code
and tests (Supplementary Fig.~\ref{fig:development}). We grouped records by
design change rather than individual commit, linking each requirement to its
implementation and subsequent checks or revision. Proposals were included as
implemented capabilities only when supported by code or other inspectable
artifacts. These retrospective examples document how requirements became
testable rules; they are separate from the formal evaluation cases.

\vspace{2pt}
\methodheading{Evaluation of discovery quality}
All evaluations used Gewu Insight commit 539d00d in isolated
databases and managed-file directories. Tasks A-C are summarized in
Supplementary Fig.~\ref{fig:evaluation-coverage}a; software and AI-rater details
are listed in Supplementary Table~\ref{tab:evaluation-reproducibility}.
Task A assessed 421 candidates: 145 from Journal Monitor, 110 from Subject
Monitor and 166 from Author Monitor. The configured sources comprised 12
journals, eight topics and 24 researchers; evaluated candidates represented
all 12 journals, six topics and 22 researchers
(Supplementary Table~\ref{tab:discovery-cohort}). Journal candidates used
1 September 2026 to 30 September 2026; subject and author candidates used
1 January 2026 to 30 September 2026. Saved source responses were replayed through
the discovery services, exercising parsing, filtering and record storage.
Candidates were sampled without replacement within each channel, with
additional random sampling within author strata, and
screened using documented document-scope and source-evidence
exclusions. Excluded items were not
replaced. Of 493 selected candidates, 421 were retained; Supplementary Table~\ref{tab:discovery-screening} details the screening counts, reasons and stages. Reported proportions describe this evaluated sample; author
inclusion probabilities differed across strata.

Journal candidates had to satisfy the specified journal, date window and
research-document type jointly. Preliminary screening did not substitute for this document-type
assessment (Supplementary Table~\ref{tab:discovery-screening}). Author attribution required the monitored
person to be an author. Strict topical relevance required the topic to be a
primary scientific question, method or result. Broad relevance additionally
included substantive methodological, application, review or adjacent-research
connections; incidental mentions were insufficient. Author attribution and
both topical-relevance measures were scored separately from date and
document-type compliance. Strict and broad relevance used the same 110
candidates, with all 78 strict-positive candidates also broad-positive.

Two AI raters independently assessed identical supplied evidence and rules
in separate contexts, without access to each other's judgments or product
predictions. Both received the same supplementary evidence and rule
clarifications for scoped reassessment; a human adjudicated the remaining
substantive disagreements. There were 14 human decisions: 12 in the base
review and two in the additional broad-topic review. A stratified sample of
candidates with two-model consensus also underwent human review
(Supplementary Table~\ref{tab:evaluation-reproducibility}).
Two author identities and one broad-topic
judgment remained unresolved and were retained in the denominators without
being counted as confirmed. Proportions in \prettyref{fig:fig3}a therefore
report confirmed candidates over all assessed candidates. Agreement between the two AI
raters was assessed before human adjudication using percentage agreement
and Cohen's $\kappa$\cite{cohen1960agreement}
(Supplementary Table~\ref{tab:evaluation-reproducibility}).

\vspace{4pt}
\methodheading{Identity and resource evaluation}
Task B comprised 160 controlled cases across ten families
(Supplementary Fig.~\ref{fig:evaluation-coverage}b). Product functions were
given specified records, file evidence and resource configurations; their
outputs and stored state were compared with the expected decisions. A case
passed only when all its applicable expectations were met. The families
covered identity, file classification, Default selection, publication notices,
annotations, preservation, validation, provenance, metadata correction and
MCP import and authorization. The two examples in \prettyref{fig:fig3}b are
independent cases from the correction-file family. Supplementary
Table~\ref{tab:task-b-cases} lists all 160 cases and their expected outcomes;
complete inputs and recorded outputs accompany the evaluation data.

\vspace{4pt}
\methodheading{State-continuity evaluation}
Task C comprised 26 scenario families, labeled C01-C26, with four distinct
scenarios per family, giving 104 in total. Supplementary
Fig.~\ref{fig:evaluation-coverage}c summarizes the families and their results;
Supplementary Table~\ref{tab:task-c-scenarios} lists the four scenarios in
each family and their expected outcomes.
The four scenarios are structural variants, not repeated runs. For example,
the four C09 scenarios remove a Library placement, an Author Monitor
association, a My Publications association or a non-Default resource,
respectively, and check that unrelated associations and retained resources
remain intact. Fig.~\ref{fig:fig3}c illustrates C09-V01.

Each scenario began with a separate database and file directory and exercised
the specified product operations. Checks compared application-visible state,
database records and filesystem inventories, including exact file hashes,
across the relevant transitions and after restart where required. A scenario
passed only when every required operation and check completed successfully.
The checks tested whether identity, resource distinctions, file contents and
paths, and unrelated contextual relations were preserved as specified for
that operation.
Tasks B and C are reported as exact pass counts without binomial confidence
intervals, because they cover deliberately constructed cases rather than
random population samples. Individual checks and supporting unit tests were
not counted as additional cases or scenarios.

\endgroup
\raggedbottom
\GewuMethodsReferences

\begingroup
\vspace{6pt}
\noindent {\bf Acknowledgments}
This work benefits from the stimulating environment fostered by the
AI Seminar Series of Westlake Physics. We thank Naizhou Wang and
Jiaxiang Cong for helpful discussions.
We also thank Long Jiao and Fazhi Yang for their assistance with
pre-release testing of Gewu Insight.
L.K. acknowledges the Tsung-Dao Lee Center for Sciences and Arts in Dali
for its hospitality during the Mingzhu Lake Conference on Quantum-AI Integration.

\noindent {\bf Author contributions}
All authors contributed to system
design, software development, evaluation, data analysis, and manuscript writing and revision. Z.W. initiated the project.

\noindent {\bf Data availability}
The data supporting the evaluation are available at
\href{https://github.com/LKong92/Gewu-Insight-Evaluation}
{\nolinkurl{github.com/LKong92/Gewu-Insight-Evaluation}}.

\noindent {\bf Code availability}
The source code for Gewu Insight is publicly available at
\href{https://github.com/LKong92/Gewu-Insight}
{\nolinkurl{github.com/LKong92/Gewu-Insight}}
under a source-available license permitting academic and non-commercial
research use. Commercial use requires prior written permission from
the copyright holders. 

\noindent {\bf Competing interests}
The authors declare no competing interests.

\endgroup

\clearpage
\begin{figure}[!htbp]
\centering
\includegraphics[width=\linewidth]{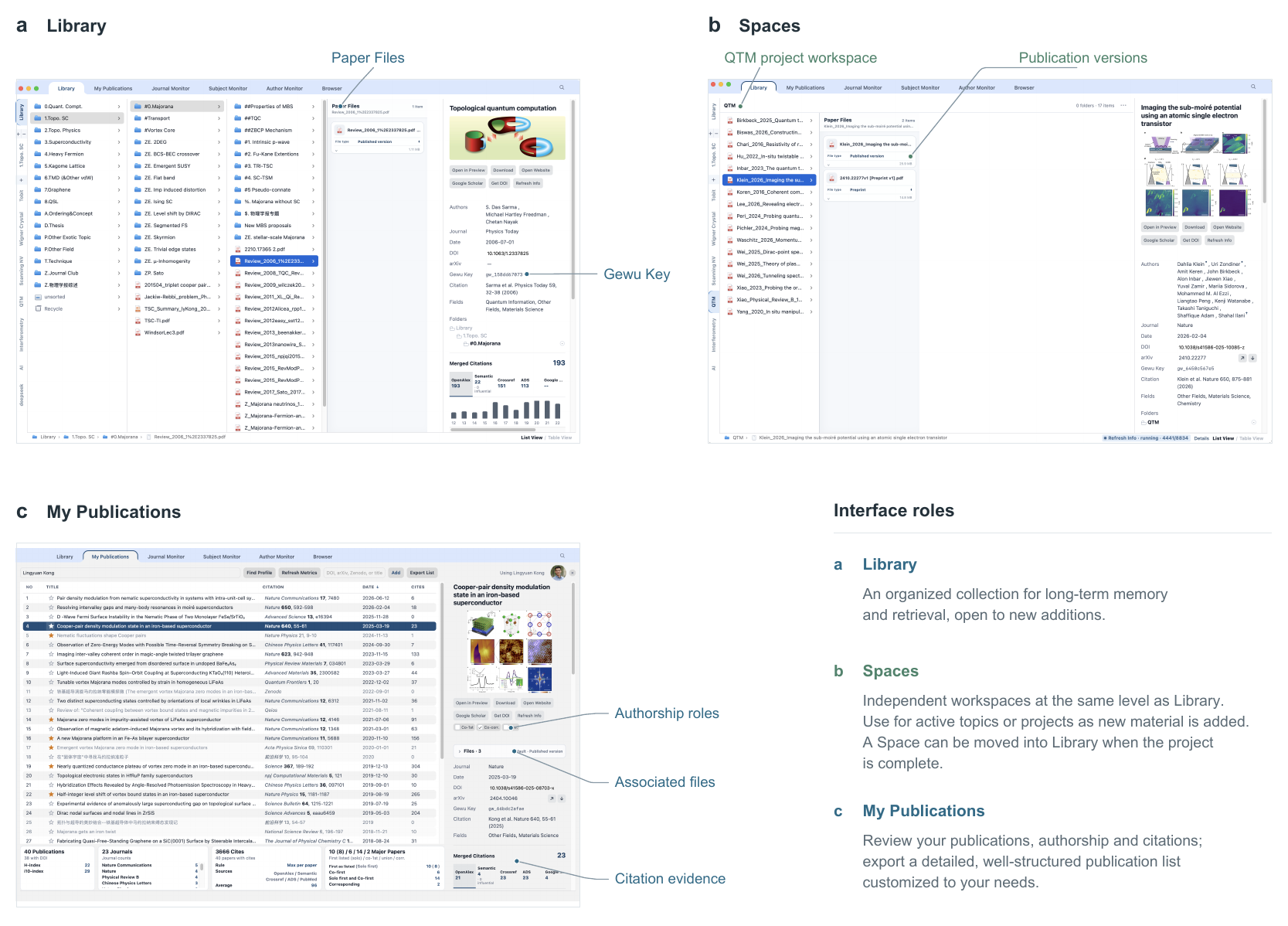}

\begingroup
\setlength{\abovecaptionskip}{3pt}
\setlength{\belowcaptionskip}{0pt}
\small
\caption{
\textbf{Personal organization and publication management in
Gewu Insight.}
\textbf{a}, Library, showing hierarchical organization, Paper Files and
metadata for the selected paper, including its Gewu Key.
\textbf{b}, An independent project Space, illustrated by the QTM workspace.
Paper Files displays a published version and a preprint associated with
the selected work.
\textbf{c}, My Publications, showing the user's publication list,
authorship roles, associated files and citation evidence. Export List
supports customized, detailed publication lists.
}
\label{fig:interface}
\endgroup

\end{figure}

\clearpage
\begin{figure}[!htbp]
\centering
\includegraphics[width=\linewidth]{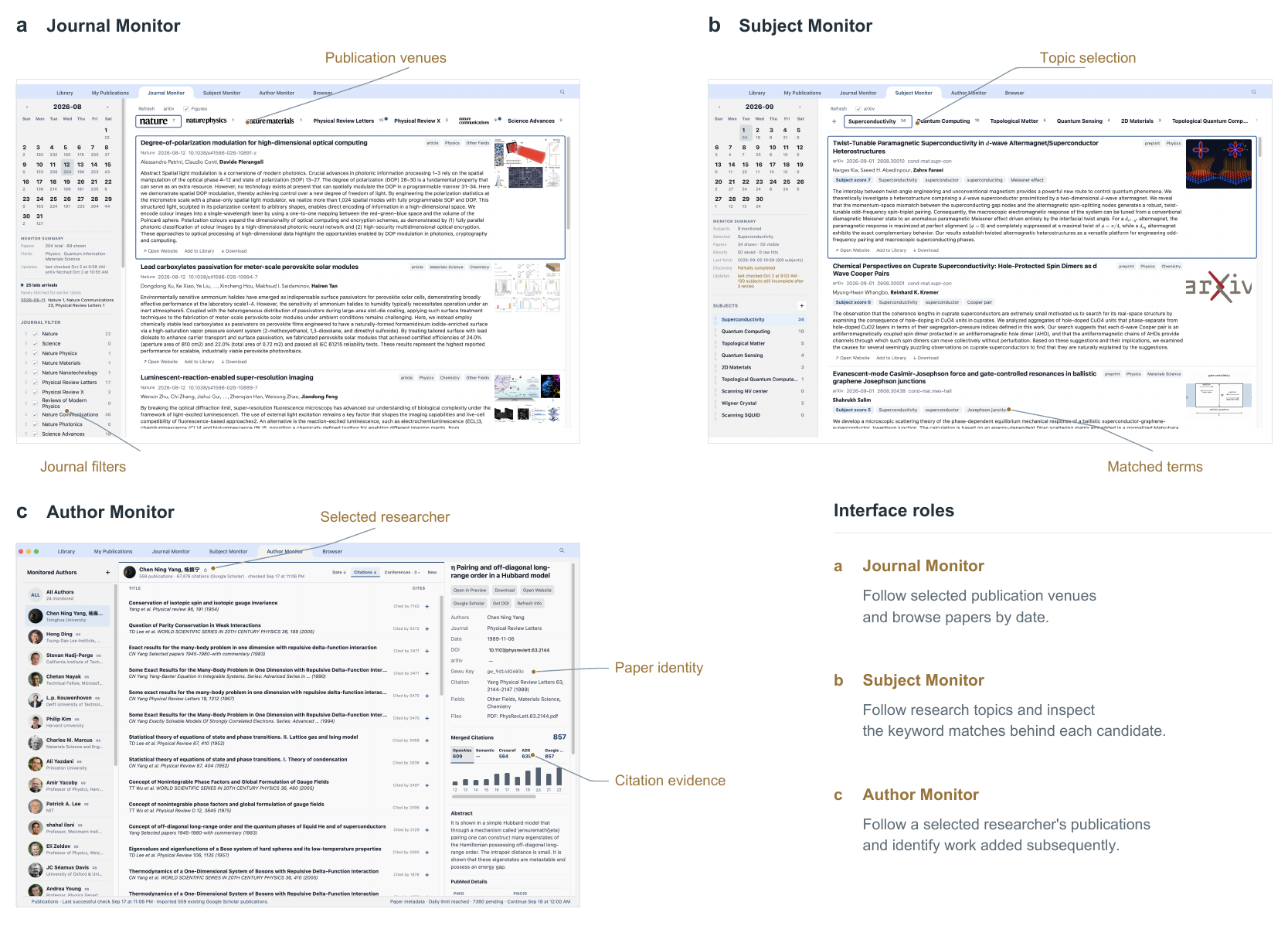}

\begingroup
\setlength{\abovecaptionskip}{3pt}
\setlength{\belowcaptionskip}{0pt}
\small
\caption{
\textbf{Complementary literature-discovery interfaces in
Gewu Insight.}
\textbf{a}, Journal Monitor, showing publication venues, calendar and
journal filters, and retrieved papers.
\textbf{b}, Subject Monitor, showing configured research topics and
candidate records with matched keywords.
\textbf{c}, Author Monitor, illustrated by Chen Ning Yang's publication
list. The detail pane displays the selected paper's identity, metadata
and source-specific citation evidence.
}
\label{fig:discovery-interface}
\endgroup

\end{figure}

\clearpage

\begin{figure}[!htbp]
\centering
\includegraphics[width=\linewidth]{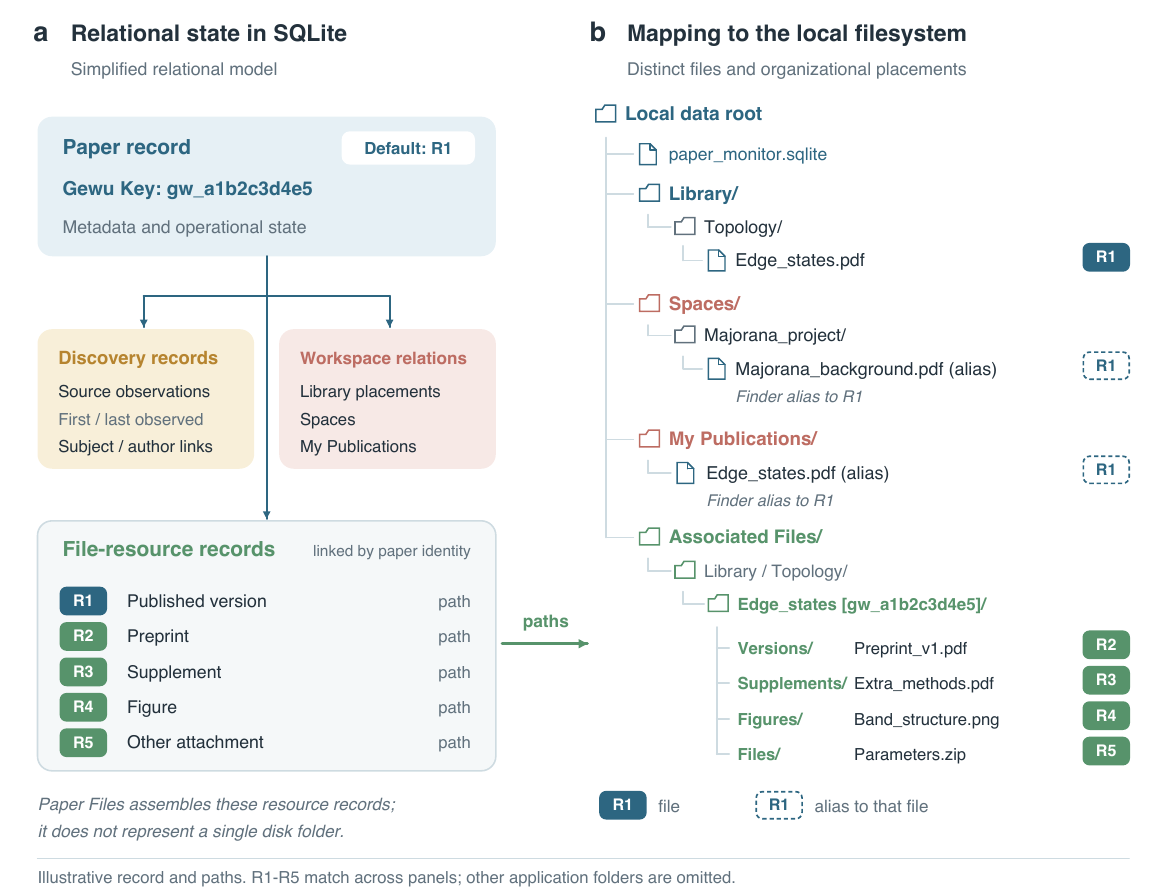}

\begingroup
\setlength{\abovecaptionskip}{3pt}
\setlength{\belowcaptionskip}{0pt}
\small
\caption{
\textbf{Relational state and its mapping to local files.}
\textbf{a}, Simplified relational model linking a canonical paper record
to discovery records, workspace relations and file resources. The Gewu Key
identifies the paper, while Default selects an associated resource.
The complete table inventory is provided in Supplementary
Table~\ref{tab:sqlite-schema}.
\textbf{b}, An illustrative directory layout for the same paper.
R1-R5 identify corresponding resource records and files across panels;
solid badges mark files and dashed badges mark filesystem links
(Finder aliases on macOS).
Additional workspace placements can expose the same primary PDF under
different names. Retained versions and associated materials are organized
under a paper-specific directory bearing the Gewu Key. Paper Files uses
database relations and stored paths to present these resources together.
Filenames and identifiers are illustrative.
}
\label{fig:storage-mapping}
\endgroup
\end{figure}

\clearpage

\begingroup
\setlength{\intextsep}{8pt plus 1pt minus 1pt}

\begin{figure}[htbp]
\centering
\includegraphics[width=\linewidth]{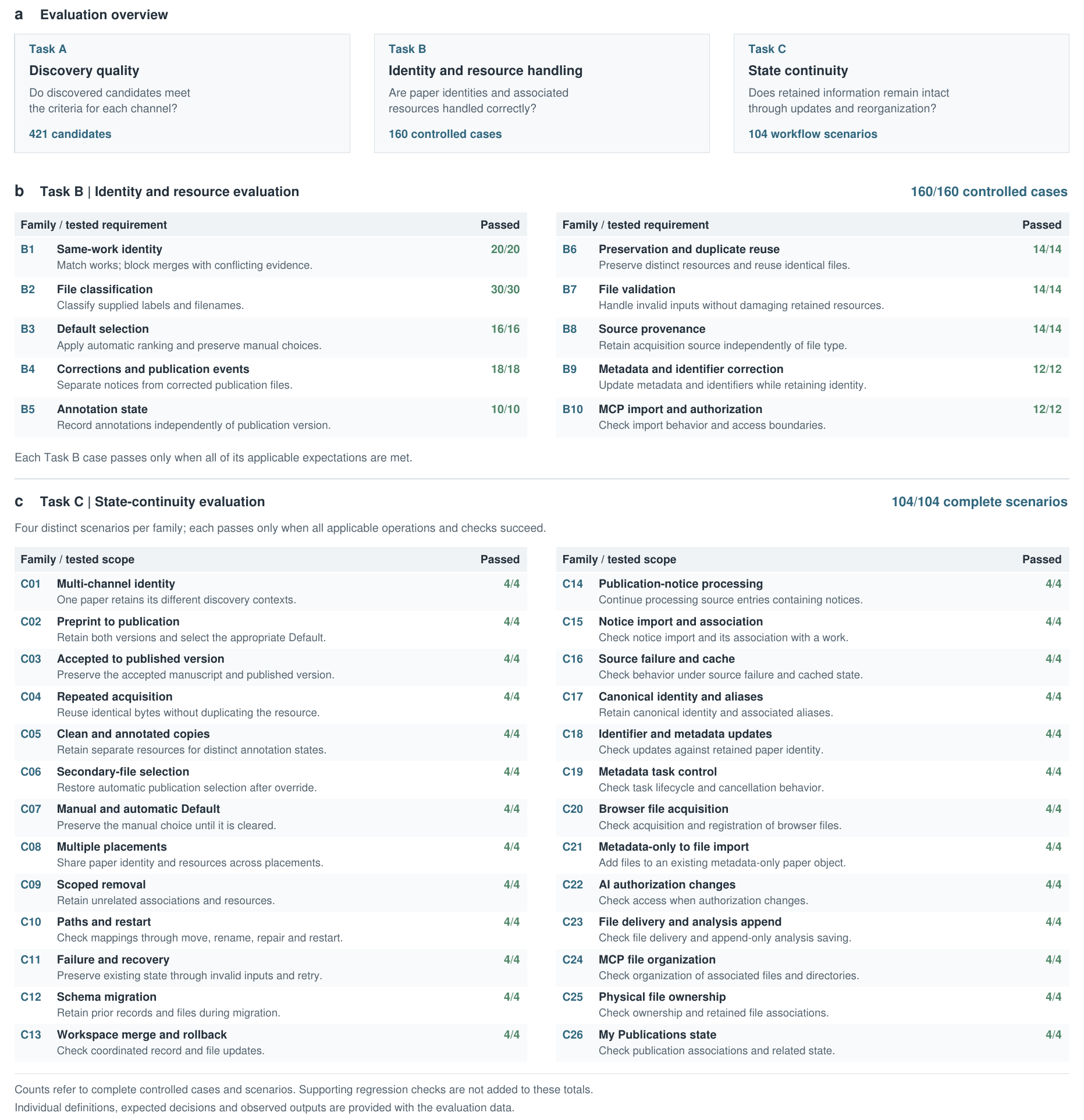}

\setlength{\abovecaptionskip}{3pt}
\setlength{\belowcaptionskip}{0pt}
\small
\caption{
\textbf{Overview and coverage of the evaluation tasks.}
\textbf{a}, Overview of Tasks A-C and their evaluated samples.
\textbf{b}, Ten Task B families and their pass counts, totaling 160 controlled
product cases. \textbf{c}, Twenty-six Task C families, each comprising four
distinct scenarios, totaling 104. Fractions count complete scenarios
satisfying every applicable check, not individual assertions or repeated
runs. Individual cases and scenarios are listed in Supplementary
Tables~\ref{tab:task-b-cases} and~\ref{tab:task-c-scenarios}; the evaluation
data provide complete definitions and outputs. The examples in
Fig.~\ref{fig:fig3}b correspond to B4-001 and B4-002; Fig.~\ref{fig:fig3}c
shows C09-V01.
}
\label{fig:evaluation-coverage}
\end{figure}

\endgroup

\clearpage
\begingroup
\setlength{\intextsep}{8pt plus 1pt minus 1pt}
\begin{figure}[h]
\centering
\includegraphics[width=\linewidth]{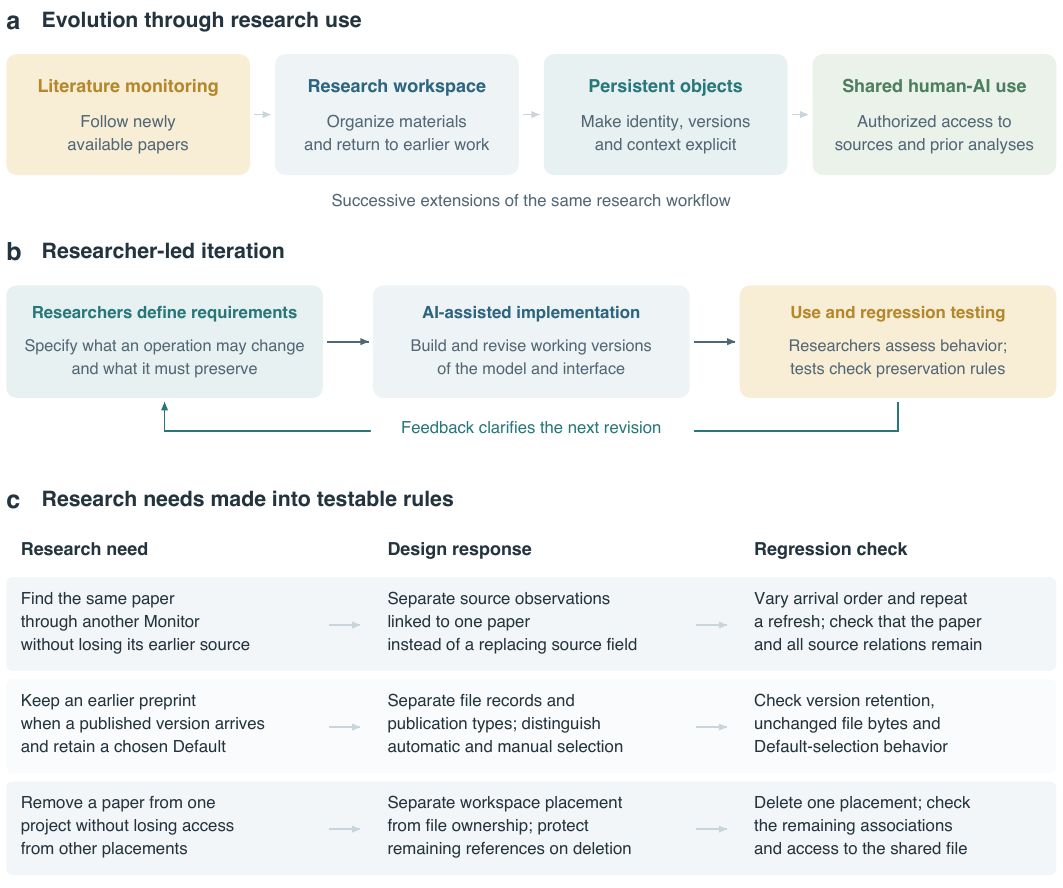}
\setlength{\abovecaptionskip}{3pt}
\setlength{\belowcaptionskip}{0pt}
\small
\caption{
\textbf{Scientist-led, AI-assisted development of Gewu Insight.}
\textbf{a}, Retrospective outline of the progression from literature monitoring
to a shared research workspace, as practical requirements became more explicit.
\textbf{b}, Researchers defined and assessed the required behavior, while
AI-assisted implementation produced working versions for further use and
testing. Feedback refined both requirements and implementation.
\textbf{c}, Three examples connect practical needs to design changes and
regression checks in the development record, showing how preservation
requirements entered the software.
}
\label{fig:development}
\end{figure}
\endgroup

\clearpage

\clearpage
\begingroup
\let\GewuCaptionFont\small
\renewcommand{\small}{\fontsize{9.5}{11.2}\selectfont}
\normalfont\rmfamily
\small
\setlength{\tabcolsep}{3pt}
\setlength{\aboverulesep}{2pt}
\setlength{\belowrulesep}{2pt}
\setlength{\lightrulewidth}{0.3pt}
\setlength{\heavyrulewidth}{0.6pt}
\setlength{\LTcapwidth}{\linewidth}
\renewcommand{\arraystretch}{1.01}
\renewcommand{\tablename}{Supplementary Table}
\makeatletter
\long\def\LT@makecaption#1#2#3{%
  \LT@mcol\LT@cols c{%
    \hbox to\z@{\hss
      \parbox[t]{\LTcapwidth}{%
        \normalfont\rmfamily\GewuCaptionFont
        #1{\textbf{#2.}\space}#3\par\vspace{4pt}%
      }\hss}%
  }%
}
\makeatother
\begin{longtable}{@{}>{\raggedright\arraybackslash}p{0.15\linewidth}
                       >{\raggedright\arraybackslash}p{0.274\linewidth}
                       >{\raggedright\arraybackslash}p{0.55\linewidth}@{}}
\caption{\textbf{SQLite tables and their roles in Gewu Insight.}
The 42 application-defined tables are grouped by function.
SQLite internal tables are excluded.}
\label{tab:sqlite-schema}\\
\toprule
\textbf{Functional group} & \textbf{Table} & \textbf{Main role} \\
\midrule
\endfirsthead
\multicolumn{3}{@{}l}{Supplementary Table~\thetable\ (continued)}\\
\toprule
\textbf{Functional group} & \textbf{Table} & \textbf{Main role} \\
\midrule
\endhead
\midrule
\multicolumn{3}{r@{}}{Continued on next page}\\
\endfoot
\bottomrule
\endlastfoot

\textbf{Identity and provenance}
& \textit{papers}
& Canonical records, Gewu Keys, metadata and Default-selection state. \\
& \textit{paper\_identifiers}
& Bibliographic identifiers linked to canonical papers. \\
& \textit{paper\_aliases}
& Earlier identifiers and Gewu Keys mapped to a surviving paper. \\
& \textit{paper\_identity\_history}
& Reconciliation events and snapshots, not a log of every edit. \\
& \textit{paper\_identity\_conflicts}
& Conflicting identity evidence and its handling status. \\
& \textit{paper\_source\_observations}
& Source relations with first- and last-observed times. \\
& \textit{publication\_notices}
& Notice identifiers, evidence and links to affected works. \\
& \textit{publication\_notice\_files}
& Files associated with publication notices. \\
\midrule

\textbf{Resources and organization}
& \textit{resources}
& File ownership, classification, provenance, paths, fingerprints and lineage relations. \\
& \textit{resource\_import\_sources}
& Import-source evidence retained when an existing file is reused. \\
& \textit{resource\_duplicate\_archive}
& Archived duplicate records, survivors and backup locations. \\
& \textit{library\_folders}
& Folder hierarchies and ordering for Library and Spaces. \\
& \textit{library\_items}
& Folder placements, display names and organization state. \\
& \textit{my\_publications}
& Personal papers, authorship roles and export flags. \\
& \textit{inventory\_import\_batches}
& Batch-import sources, configuration, status and summary results. \\
\midrule

\textbf{Acquisition and Journal}
& \textit{fetch\_runs}
& Source, time window, status and counts for acquisition runs. \\
& \textit{journal\_subscriptions}
& Journal subscriptions, source identifiers and retrieval settings. \\
& \textit{journal\_monitor\_papers}
& Paper associations with Journal Monitor runs and observed sources. \\
\midrule

\textbf{Subject discovery}
& \textit{theme\_results}
& Subject matches, scores, keyword hits and search windows. \\
& \textit{theme\_source\_runs}
& Source-specific acquisition runs, outcomes and retry state. \\
& \textit{theme\_source\_tasks}
& Individual queries and their outcomes within an acquisition run. \\
& \textit{theme\_candidates}
& Candidate papers linked to sources, time windows and query plans. \\
& \textit{subject\_preserved\_windows}
& Search windows protected from rematching after subject edits. \\
& \textit{aps\_rss\_feeds}
& APS RSS feed state and cache-validation metadata. \\
& \textit{aps\_rss\_entries}
& Feed-to-paper relations, publication dates and observation times. \\
\midrule

\textbf{Author discovery}
& \textit{author\_monitors}
& Researcher profiles, source identities and monitoring state. \\
& \textit{author\_monitor\_papers}
& Researcher-paper relations, observation times and unread state. \\
& \textit{author\_monitor\_card\_cache}
& Cached paper-card payloads for Author Monitor display. \\
& \textit{author\_monitor\_card\_cache\_dirty}
& Revision markers identifying cached cards that need updating. \\
\midrule

\textbf{Operations and settings}
& \textit{paper\_reads}
& Seen-time markers for late-arriving papers, not reading notes. \\
& \textit{paper\_metadata\_refreshes}
& Metadata-refresh attempts, modes, timings, outcomes and errors. \\
& \textit{publication\_import\_jobs}
& Resumable My Publications import jobs, checkpoints and progress. \\
& \textit{publication\_import\_items}
& Per-item import outcomes and links to canonical papers. \\
& \textit{publication\_identifier\_tasks}
& Identifier-resolution tasks and retry state for My Publications entries. \\
& \textit{app\_settings}
& Application configuration and other persisted key-value state. \\
\midrule

\textbf{Browser}
& \textit{browser\_bookmarks}
& Browser bookmarks, bookmark folders and their ordering. \\
& \textit{browser\_history}
& Browsing-history entries, separate from canonical paper records. \\
\midrule

\textbf{AI access}
& \textit{ai\_connections}
& Connection grants, token hashes, collection scope, permissions and expiry. \\
& \textit{ai\_audit}
& AI-operation logs and outcomes for each connection. \\
& \textit{ai\_analyses}
& Analysis-save receipts and source evidence; text is stored in files. \\
& \textit{ai\_write\_requests}
& Request IDs and saved results for idempotent AI writes. \\
& \textit{ai\_uploads}
& Staged upload metadata, file hashes and processing status. \\

\end{longtable}

\clearpage
\begingroup
\renewcommand{\small}{\fontsize{9.5}{11.2}\selectfont}
\small
\renewcommand{\arraystretch}{1.0}
\begin{longtable}{@{}>{\raggedright\arraybackslash}p{0.61\linewidth}
                       >{\raggedleft\arraybackslash}p{0.10\linewidth}
                       >{\raggedleft\arraybackslash}p{0.12\linewidth}
                       >{\raggedleft\arraybackslash}p{0.11\linewidth}@{}}
\caption{\textbf{Discovery evaluation cohort and sample composition.}
Assessed and confirmed counts cover all configured monitors. Confirmation
denotes joint journal/date/document-type eligibility, strict topical relevance
or target-author attribution, respectively. Broad counts include strict-positive
subject candidates. Unresolved judgments (two author, one broad-topic) remain
in assessed counts. Zero indicates no evaluated sample. Author identifiers
follow the evaluation data; see Methods for criteria and sampling.}
\label{tab:discovery-cohort}\\
\noalign{\vskip 8pt}
\toprule
\textbf{Monitor or topic} & \textbf{Assessed} & \textbf{Confirmed} & \textbf{Broad} \\
\midrule
\endfirsthead
\multicolumn{4}{@{}l}{Supplementary Table~\thetable\ (continued)}\\
\toprule
\textbf{Monitor or topic} & \textbf{Assessed} & \textbf{Confirmed} & \textbf{Broad} \\
\midrule
\endhead
\midrule
\multicolumn{4}{r@{}}{Continued on next page}\\
\endfoot
\bottomrule
\endlastfoot
\textbf{Journal Monitor} & \textbf{145} & \textbf{137} & -- \\
Nature & 9 & 9 & -- \\
Science & 7 & 2 & -- \\
Nature Physics & 2 & 1 & -- \\
Nature Materials & 1 & 1 & -- \\
Nature Nanotechnology & 1 & 1 & -- \\
Nature Photonics & 2 & 2 & -- \\
Nature Communications & 58 & 58 & -- \\
Science Advances & 18 & 17 & -- \\
Physical Review Letters & 21 & 21 & -- \\
Physical Review X & 3 & 3 & -- \\
Reviews of Modern Physics & 2 & 2 & -- \\
Nano Letters & 21 & 20 & -- \\
\midrule
\textbf{Subject Monitor} & \textbf{110} & \textbf{78} & \textbf{104} \\
Two-dimensional magnetism & 16 & 16 & 16 \\
Nonequilibrium quantum dynamics & 9 & 7 & 9 \\
Perovskite photovoltaic stability & 6 & 6 & 6 \\
Solid-state battery interfaces & 39 & 29 & 39 \\
Machine learning in catalysis & 0 & -- & -- \\
Gravitational-wave data analysis & 0 & -- & -- \\
Protein structure prediction & 32 & 13 & 27 \\
Ocean carbon cycle & 8 & 7 & 7 \\
\midrule
\textbf{Author Monitor} & \textbf{166} & \textbf{156} & -- \\
AUTHOR-01 & 15 & 9 & -- \\
AUTHOR-02 & 5 & 5 & -- \\
AUTHOR-03 & 10 & 10 & -- \\
AUTHOR-04 & 7 & 7 & -- \\
AUTHOR-05 & 2 & 2 & -- \\
AUTHOR-06 & 0 & -- & -- \\
AUTHOR-07 & 10 & 10 & -- \\
AUTHOR-08 & 17 & 17 & -- \\
AUTHOR-09 & 9 & 9 & -- \\
AUTHOR-10 & 0 & -- & -- \\
AUTHOR-11 & 6 & 6 & -- \\
AUTHOR-12 & 14 & 13 & -- \\
AUTHOR-13 & 2 & 1 & -- \\
AUTHOR-14 & 7 & 7 & -- \\
AUTHOR-15 & 8 & 8 & -- \\
AUTHOR-16 & 6 & 6 & -- \\
AUTHOR-17 & 9 & 9 & -- \\
AUTHOR-18 & 7 & 7 & -- \\
AUTHOR-19 & 4 & 3 & -- \\
AUTHOR-20 & 2 & 2 & -- \\
AUTHOR-21 & 9 & 9 & -- \\
AUTHOR-22 & 6 & 5 & -- \\
AUTHOR-23 & 4 & 4 & -- \\
AUTHOR-24 & 7 & 7 & -- \\
\end{longtable}
\clearpage
{\fontsize{10}{12}\selectfont
\refstepcounter{table}\label{tab:discovery-screening}
{\normalfont\rmfamily\GewuCaptionFont\noindent\textbf{Supplementary Table~\thetable. Screening of selected discovery candidates.}\par}
\medskip
\noindent\begin{tabular*}{\linewidth}{@{\extracolsep{\fill}}l r r r r@{}}
\toprule
\textbf{Channel} & \textbf{Selected} & \shortstack{\textbf{Excluded:}\\\textbf{document scope}} & \shortstack{\textbf{Excluded:}\\\textbf{unverifiable source}} & \textbf{Evaluated} \\
\midrule
Journal & 150 & 5 & 0 & 145 \\
Subject & 110 & 0 & 0 & 110 \\
Author & 233 & 49 & 18 & 166 \\
\midrule
\textbf{Total} & \textbf{493} & \textbf{54} & \textbf{18} & \textbf{421} \\
\bottomrule
\end{tabular*}\par
\medskip
\noindent Screening was applied to sampled software outputs before final scoring.
The five Journal exclusions comprised two editorial or curated-collection
items and three news, commentary or digest items.
These five exclusions were recorded during preliminary material
preparation. Among the 145 retained Journal candidates, eight were subsequently
judged not to be research documents and remained in the denominator as failures.
Thus, 137/145 is conditional on preliminary screening, rather than a precision
estimate for all 150 sampled outputs. The 49 Author document-scope
exclusions included conference abstracts, posters or programmes, theses,
patents and other out-of-scope documents. Complete conference papers remained
eligible. The screening ledger includes one conference abstract excluded
during review under the shared document-scope rule.\par
\medskip
\noindent Evidence collection respected publisher access policies. Eighteen Author
candidates were excluded because their exact sources could not be located
or verified from accessible material; these were not confirmed document-type
exclusions. Excluded candidates were not replaced or counted as passes.
Retained candidates with unresolved judgments remained in the denominators.
Counts refer to monitor-work candidates, with each excluded candidate counted
once; detailed records accompany the evaluation data.}
\endgroup

\clearpage
\renewcommand{\small}{\fontsize{9.5}{11.2}\selectfont}
\small
\setlength{\aboverulesep}{1pt}
\setlength{\belowrulesep}{1pt}
\setlength{\LTpost}{4pt}
\begin{longtable}{@{}>{\raggedright\arraybackslash}p{0.14\linewidth}
                       >{\raggedright\arraybackslash}p{0.25\linewidth}
                       >{\raggedright\arraybackslash}p{0.17\linewidth}
                       >{\raggedright\arraybackslash}p{0.40\linewidth}@{}}
\caption{\textbf{MCP tools, permissions and operation boundaries.}
All 24 tools are grouped by function. Permissions are collection-scoped;
paired permissions follow tool order. Tool discovery follows the current
grant, except for local read-copy management. Gewu Keys identify papers,
not access rights.}
\label{tab:mcp-capabilities}\\
\toprule
\textbf{Function} & \textbf{MCP tool(s)} & \textbf{Permission} & \textbf{Operation and preservation boundary} \\
\midrule
\endfirsthead
\multicolumn{4}{@{}l}{Supplementary Table~\thetable\ (continued)}\\
\toprule
\textbf{Function} & \textbf{MCP tool(s)} & \textbf{Permission} & \textbf{Operation and preservation boundary} \\
\midrule
\endhead
\midrule
\multicolumn{4}{r@{}}{Continued on next page}\\
\endfoot
\bottomrule
\endlastfoot

\textbf{Paper lookup}
& \textit{search\_papers}\newline \textit{get\_paper}
& \textit{papers:read}
& Search saved metadata; retrieve a paper's metadata, registered files and
Default. No external literature search. \\
\midrule
\textbf{Collection browsing}
& \textit{list\_folders}\newline \textit{list\_folder\_papers}
& \textit{folders:read}
& Browse authorized Library/Space folders and paper placements. Repeated
placements retain the same Gewu Key. \\
\midrule
\textbf{File reading}
& \textit{read\_paper}\newline \textit{get\_paper\_file}
& \textit{files:read}
& Extract PDF pages or text chunks, or deliver a read-only client copy with
file ID and SHA-256. Originals remain unchanged. \\
\midrule
\textbf{Read-copy management}
& \textit{get\_file\_delivery\_status}\newline \textit{release\_paper\_file}
& Local session only
& Inspect temporary-copy quota; release a delivered copy. Remains available
after revocation, without retrieving or deleting managed files. \\
\midrule
\textbf{Analysis saving}
& \textit{save\_analysis}
& \textit{analysis:write};
\textit{files:read} for source-file evidence
& Append Markdown to Paper Files. Preserve earlier notes, source files and
any existing Default selection; check the supplied source hash before saving. \\
\midrule
\textbf{Workspace creation}
& \textit{create\_space}\newline \textit{create\_folder}
& \textit{spaces:create}\newline \textit{folders:create}
& Create a Space or child folder. The new Space becomes writable by this
connection; existing destinations require a write grant. \\
\midrule
\textbf{Upload staging}
& \textit{upload\_file}\newline \textit{get\_upload\_status}\newline \textit{release\_upload}
& Any of \textit{papers:import}, \textit{files:import}, \textit{files:add}
& Stage a client file, inspect quota or release unused staging. Registration
is separate; successful import frees that file's staging. \\
\midrule
\textbf{Paper import}
& \textit{import\_paper}
& \textit{papers:import}
& Import supplied paper metadata, optionally with a PDF. Matching DOI/arXiv evidence reuses
an authorized paper; conflicts block import. Preserve existing files and
manual Default; automatic Default follows version priority. \\
\midrule
\textbf{Other resources}
& \textit{import\_file}\newline \textit{add\_paper\_file}
& \textit{files:import}\newline \textit{files:add}
& Import an independent non-PDF file, or attach a file to an existing paper.
Exact bytes reuse resources. Preserve content and manual Default;
automatic Default follows version priority. \\
\midrule
\textbf{Metadata completion}
& \textit{update\_paper\_metadata}
& \textit{metadata:write}
& Fill missing fields from supplied facts and record disagreements.
Preserve existing values; no online lookup. \\
\midrule
\textbf{Metadata lookup}
& \textit{resolve\_paper}\newline \textit{get\_resolve\_status}
& \textit{papers:resolve};
\textit{files:read} if PDF reading is needed
& Start or inspect a separately requested lookup for an existing paper,
not an import step. Preserve existing values, files and Default. \\
\midrule
\textbf{Folder organization}
& \textit{move\_folder}\newline \textit{rename\_folder}
& \textit{folders:move}\newline \textit{folders:rename}
& Move a subtree or rename a folder/Space, synchronizing local paths and
links while preserving identities, file bytes and Default. \\
\midrule
\textbf{File organization}
& \textit{move\_file}\newline \textit{rename\_file}
& \textit{files:move}\newline \textit{files:rename}
& Move or rename placements/files; transfer ownership. Preserve file IDs,
bytes and classification. Transfers can reselect the source Default or
retire an empty source record. \\
\end{longtable}
\noindent Import, metadata and organization require writable Library/Space
collections; moves/renames check all affected locations. Analysis saving is
separately granted. Request IDs support exact write retries. No tool edits
existing file contents or directly deletes managed resources. Grants neither
restrict independent filesystem access nor recall delivered copies.

\clearpage
\renewcommand{\small}{\fontsize{10}{12}\selectfont}
\small
\setlength{\aboverulesep}{4pt}
\setlength{\belowrulesep}{4pt}
\begin{longtable}{@{}>{\raggedright\arraybackslash}p{0.23\linewidth}
                       >{\raggedright\arraybackslash}p{0.75\linewidth}@{}}
\caption{\textbf{Software snapshot and AI-reference construction.}
Model names identify the configurations used in the evaluation; unrecorded
backend versions or settings are not inferred.}
\label{tab:evaluation-reproducibility}\\
\toprule
\textbf{Item} & \textbf{Specification} \\
\midrule
\endfirsthead
\multicolumn{2}{@{}l}{Supplementary Table~\thetable\ (continued)}\\
\toprule
\textbf{Item} & \textbf{Specification} \\
\midrule
\endhead
\midrule
\multicolumn{2}{r@{}}{Continued on next page}\\
\endfoot
\bottomrule
\endlastfoot
\textbf{Software and environment}
& Gewu Insight commit
539d00d,
database schema 22; macOS and CPython 3.9.6. Tasks A-C used the same
revision and isolated databases and file directories. Task A replayed
saved source responses. \\
\midrule
\textbf{AI reviewers}
& Codex (GPT-6 Astra) and DeepSeek Harness (DeepSeek V4.1 Flash; Max configuration)
reviewed Task A independently. \\
\midrule
\textbf{Evidence and instructions}
& Both reviewers received identical case evidence and rules, without
product predictions or each other's judgments. Shared evidence supplements
and rule clarifications preceded scoped reassessment. Instructions,
accepted judgments and evidence locators are retained in the evaluation data. \\
\midrule
\textbf{Agreement before adjudication}
& The base review covered 1,263 fields, three per discovery candidate.
Initial agreement was 1,225/1,263 (96.99\%; $\kappa=0.8519$),
rising to 1,251/1,263 (99.05\%; $\kappa=0.9534$) after
supplementation and corrections.
For the separate 110-field broad-topic review, agreement was initially
100/110 (90.91\%) and subsequently 108/110 (98.18\%; $\kappa=0.8487$),
including one shared uncertain judgment. These are inter-model agreement
measures, not estimates of correctness. \\
\midrule
\textbf{Human adjudication and review}
& A human supplied 14 decisions on residual disagreements (12 base and two
broad-topic).
A separate human review of a stratified sample of 30 consensus
candidates without prior human adjudication covered 10 per channel: three
fields per journal candidate, two per subject candidate and author attribution.
Final human judgments agreed with the two-model consensus in all 60 comparisons.
Two author identities and one broad-topic judgment remain unresolved. The
reference combines dual-model consensus and selected human adjudication. \\
\midrule
\textbf{Sampling and reporting scope}
& Screening applied the documented document-scope and source-evidence
exclusions; excluded items were not replaced. Author sampling included
within-author supplementation with unequal inclusion probabilities.
The reported 156/166 is an unweighted evaluated-sample proportion;
the inverse-probability-weighted confirmed proportion is 91.93\%. Weights reflect the recorded author sampling strata within the screened evaluation scope.
Unknown labels remain in denominators and are not counted as confirmed. \\
\midrule
\textbf{Reporting definitions}
& Broad relevance was added as a supplementary assessment of the same 110
candidates after inspection of the initial results; strict-relevance labels
were unchanged. \\
\midrule
\textbf{Strict and broad relevance: example}
& For ``Nonequilibrium quantum dynamics'', the strict definition specified
quantum many-body systems. A periodically driven Ising-chain study\cite{gadge2026temporal}
qualified under both definitions because its central question concerned many-body
entanglement dynamics. A driven-dissipative Landau-polariton study\cite{mivehvar2026driven}
qualified only under the broad definition: it directly analyzed nonequilibrium
evolution in a single-particle cavity system. Broad relevance therefore
included related physical systems outside the original many-body boundary;
the strict labels were retained. \\
\midrule
\textbf{Reference and release records}
& Final labels retain their consensus or human-adjudication origin.
Protocols, case definitions, reviewer judgments, scored outputs and
recalculation scripts accompany the evaluation data. Source locators and
hashes identify supporting evidence; original literature files remain in
the separate evidence archive. \\
\end{longtable}

\clearpage
\renewcommand{\small}{\fontsize{10}{12}\selectfont}
\small
\setlength{\tabcolsep}{4pt}
\setlength{\aboverulesep}{2pt}
\setlength{\belowrulesep}{2pt}
\renewcommand{\arraystretch}{1.04}
\begin{longtable}{@{}>{\raggedright\arraybackslash}p{0.12\linewidth}
                       >{\raggedright\arraybackslash}p{0.43\linewidth}
                       >{\raggedright\arraybackslash}p{\dimexpr0.45\linewidth-16pt\relax}@{}}
\caption{\textbf{Individual identity and resource evaluation cases (Task B).}
All 160 controlled cases passed. The rows summarize the supplied conditions
and expected outcomes; complete inputs and recorded outputs accompany the
evaluation data. Cases used synthetic records and files; B2 tests
classification from supplied labels and filenames.}
\label{tab:task-b-cases}\\
\caseboundaryrule
\textbf{ID} & \textbf{Input or operation} & \textbf{Expected outcome} \\
\caseboundaryrule
\endfirsthead
\multicolumn{3}{@{}l}{Supplementary Table~\thetable\ (continued)}\\
\caseboundaryrule
\textbf{ID} & \textbf{Input or operation} & \textbf{Expected outcome} \\
\caseboundaryrule
\endhead
\caseboundaryrule
\multicolumn{3}{r@{}}{Continued on next page}\\
\endfoot
\endlastfoot
\casefamilyfirst{B1\quad Same-work identity (20 cases)}
  {B1-001}
  {Ingest a second observation with matching metadata.}
  {Accept it for the existing work.}
\mbox{B1-002} & Add a compatible observation with an extended author list. & Accept it for the existing work. \\[2pt]
\mbox{B1-003} & Add a compatible title ending in ``accepted manuscript''. & Accept it for the existing work. \\[2pt]
\mbox{B1-004} & Use the journal abbreviation Phys. Rev. B. & Accept the compatible observation. \\[2pt]
\mbox{B1-005} & Supply a DOI landing URL matching the recorded DOI. & Accept the compatible observation. \\[2pt]
\mbox{B1-006} & Supply a non-DOI publisher landing URL. & Accept the otherwise compatible observation. \\[2pt]
\mbox{B1-007} & Reuse the DOI with conflicting biology metadata. & Reject the conflicting observation. \\[2pt]
\mbox{B1-008} & Reuse the DOI with conflicting climate metadata. & Reject the conflicting observation. \\[2pt]
\mbox{B1-009} & Reuse the DOI with conflicting clinical metadata. & Reject the conflicting observation. \\[2pt]
\mbox{B1-010} & Reuse the DOI with conflicting astronomy metadata. & Reject the conflicting observation. \\[2pt]
\mbox{B1-011} & Supply a conflicting DOI URL under the same DOI prefix. & Reject the conflicting observation. \\[2pt]
\mbox{B1-012} & Supply a conflicting DOI URL under another DOI prefix. & Reject the conflicting observation. \\[2pt]
\mbox{B1-013} & Ingest two provider records carrying the same DOI. & Retain one canonical paper record. \\[2pt]
\mbox{B1-014} & Supply the same DOI as an uppercase DOI URL. & Normalize the identifier and retain one paper. \\[2pt]
\mbox{B1-015} & Ingest two provider records with the same arXiv identifier. & Retain one canonical paper record. \\[2pt]
\mbox{B1-016} & Ingest equal titles with different DOIs. & Keep two separate paper records. \\[2pt]
\mbox{B1-017} & Ingest equal titles without identifiers. & Keep two separate paper records. \\[2pt]
\mbox{B1-018} & Add a DOI that connects a second record to an existing work. & Retain one paper; its earlier Gewu Key still resolves. \\[2pt]
\mbox{B1-019} & Write the same DOI concurrently through two connections. & Retain one canonical paper record. \\[2pt]
\mbox{B1-020} & Reuse an existing DOI with conflicting provider metadata. & Do not create or merge in the conflicting record. \\[2pt]
\midrule
\casefamilyfirst{B2\quad File classification (30 cases)}
  {B2-001}
  {Supply publication-version role and preprint label.}
  {Return preprint.}
\mbox{B2-002} & Supply publication-version role and accepted-manuscript label. & Return accepted manuscript. \\[2pt]
\mbox{B2-003} & Supply publication-version role and version-of-record label. & Return published version. \\[2pt]
\mbox{B2-004} & Supply publication-version role and first-published label. & Return published version. \\[2pt]
\mbox{B2-005} & Supply publication-version role and corrected-published label. & Return corrected published version. \\[2pt]
\mbox{B2-006} & Supply other role and filename ``Author Correction.pdf''. & Return correction notice. \\[2pt]
\mbox{B2-007} & Supply other role and filename ``Publisher Corrigendum.pdf''. & Return correction notice. \\[2pt]
\mbox{B2-008} & Supply supplement role and filename ``supporting information.pdf''. & Return supplementary material. \\[2pt]
\mbox{B2-009} & Supply peer-review role and filename ``peer review file.pdf''. & Return peer-review file. \\[2pt]
\mbox{B2-010} & Supply other role and filename ``decision letter.pdf''. & Return peer-review file. \\[2pt]
\mbox{B2-011} & Supply data role and filename ``dataset.csv''. & Return other. \\[2pt]
\mbox{B2-012} & Supply a readable PDF without a role or version label. & Return unknown. \\[2pt]
\mbox{B2-013} & Supply publication-version role but an invalid parser status. & Return invalid file. \\[2pt]
\mbox{B2-014} & Supply version-of-record label for an article with appended supplement. & Return published version. \\[2pt]
\mbox{B2-015} & Supply accepted-manuscript label for ``Nature Accelerated Article Preview.pdf''. & Return accepted manuscript. \\[2pt]
\mbox{B2-016} & Supply accepted-manuscript label for ``Science First Release.pdf''. & Return accepted manuscript. \\[2pt]
\mbox{B2-017} & Supply version-of-record label for a publisher article with appendix. & Return published version. \\[2pt]
\mbox{B2-018} & Supply corrected-version label for ``corrected article.pdf''. & Return corrected published version. \\[2pt]
\mbox{B2-019} & Supply other role and filename ``reviewer report 2.pdf''. & Return peer-review file. \\[2pt]
\mbox{B2-020} & Supply note role and filename ``lab notes.txt''. & Return other. \\[2pt]
\mbox{B2-021} & Supply other role and a filename beginning ``Retraction''. & Return retraction notice. \\[2pt]
\mbox{B2-022} & Supply other role and a filename beginning ``Expression of concern''. & Return expression of concern. \\[2pt]
\mbox{B2-023} & Supply other role and a filename beginning ``Withdrawal''. & Return withdrawal notice. \\[2pt]
\mbox{B2-024} & Supply other role and a filename beginning ``Addendum''. & Return addendum. \\[2pt]
\mbox{B2-025} & Normalize an explicit perspective classification. & Return perspective. \\[2pt]
\mbox{B2-026} & Normalize an explicit News and Views classification. & Return News and Views. \\[2pt]
\mbox{B2-027} & Normalize an explicit summary classification. & Return summary. \\[2pt]
\mbox{B2-028} & Normalize an explicit analysis classification. & Return analysis. \\[2pt]
\mbox{B2-029} & Normalize an explicit source-data classification. & Return source data. \\[2pt]
\mbox{B2-030} & Normalize an explicit source-code classification. & Return source code. \\[2pt]
\midrule
\casefamilyfirst{B3\quad Default selection (16 cases)}
  {B3-001}
  {Register a preprint only.}
  {Select the preprint as Default.}
\mbox{B3-002} & Register a preprint and accepted manuscript. & Select the accepted manuscript as Default. \\[2pt]
\mbox{B3-003} & Register a preprint, accepted manuscript and published version. & Select the published version as Default. \\[2pt]
\mbox{B3-004} & Register published and corrected published versions. & Select the corrected published version as Default. \\[2pt]
\mbox{B3-005} & Register a published version and correction notice. & Keep the published version as Default. \\[2pt]
\mbox{B3-006} & Register supplementary material and a peer-review file only. & Select neither as an automatic publication Default. \\[2pt]
\mbox{B3-007} & Register a published version, supplement and correction notice. & Select the published version as Default. \\[2pt]
\mbox{B3-008} & Register a preprint and a file marked invalid. & Select the preprint as Default. \\[2pt]
\mbox{B3-009} & Manually select a preprint before adding published and corrected versions. & Preserve the manual preprint selection. \\[2pt]
\mbox{B3-010} & Register a published version and retraction notice. & Keep the published version as Default. \\[2pt]
\mbox{B3-011} & Register a published version and expression of concern. & Keep the published version as Default. \\[2pt]
\mbox{B3-012} & Register a preprint and withdrawal notice. & Keep the preprint as Default. \\[2pt]
\mbox{B3-013} & Register a published version and addendum. & Keep the published version as Default. \\[2pt]
\mbox{B3-014} & Register source data, source code and an analysis only. & Select none as an automatic publication Default. \\[2pt]
\mbox{B3-015} & Register an unknown file and a published version. & Select the published version as Default. \\[2pt]
\mbox{B3-016} & Clear a manual preprint selection after adding a published version. & Restore automatic selection of the published version. \\[2pt]
\midrule
\casefamilyfirst{B4\quad Corrections and publication events (18 cases)}
  {B4-001}
  {Register a published version and a correction notice.}
  {Retain both files; the published version is Default.}
\mbox{B4-002} & Register published and corrected versions plus a correction notice. & Retain all three files; the corrected version is Default. \\[2pt]
\mbox{B4-003} & Register a correction notice only. & Retain the notice without a publication Default. \\[2pt]
\mbox{B4-004} & Register a preprint and correction notice. & Retain both files; the preprint is Default. \\[2pt]
\mbox{B4-005} & Register an accepted manuscript and correction notice. & Retain both files; the accepted manuscript is Default. \\[2pt]
\mbox{B4-006} & Register published and corrected published versions. & Retain both files; the corrected version is Default. \\[2pt]
\mbox{B4-007} & Ingest a correction notice with one explicit target DOI. & Store one linked correction event, not another paper. \\[2pt]
\mbox{B4-008} & Ingest a retraction notice with one explicit target DOI. & Store one linked retraction event, not another paper. \\[2pt]
\mbox{B4-009} & Ingest an expression of concern with one explicit target DOI. & Store one linked concern event, not another paper. \\[2pt]
\mbox{B4-010} & Ingest a withdrawal notice with one explicit target DOI. & Store one linked withdrawal event, not another paper. \\[2pt]
\mbox{B4-011} & Ingest an addendum with one explicit target DOI. & Store one linked addendum event, not another paper. \\[2pt]
\mbox{B4-012} & Ingest a correction notice without a target DOI. & Retain an event awaiting association; create no paper. \\[2pt]
\mbox{B4-013} & Ingest a correction notice with multiple target DOIs. & Retain an event awaiting review; do not choose a target. \\[2pt]
\mbox{B4-014} & Ingest a notice whose target DOI is its own DOI. & Leave the event unlinked; do not associate it with itself. \\[2pt]
\mbox{B4-015} & Ingest the same linked correction notice twice. & Retain one event and one original paper. \\[2pt]
\mbox{B4-016} & Inspect an original article carrying a retracted-status flag. & Do not classify the original article as a notice. \\[2pt]
\mbox{B4-017} & Inspect a full article marked as a corrected published version. & Do not classify the article as a notice. \\[2pt]
\mbox{B4-018} & Change the target relation of an already recorded notice. & Require review and preserve the existing event. \\[2pt]
\midrule
\casefamilyfirst{B5\quad Annotation state (10 cases)}
  {B5-001}
  {Attach a clean one-page PDF.}
  {Record it as unannotated.}
\mbox{B5-002} & Attach a PDF containing a text annotation. & Record it as annotated. \\[2pt]
\mbox{B5-003} & Attach a clean two-page PDF. & Record it as unannotated. \\[2pt]
\mbox{B5-004} & Override a text-annotated PDF as unannotated; refresh fingerprints. & Preserve the manual unannotated state. \\[2pt]
\mbox{B5-005} & Attach clean and text-annotated copies to one paper. & Retain two separate resources. \\[2pt]
\mbox{B5-006} & Attach clean PDFs with different page counts. & Retain two separate resources. \\[2pt]
\mbox{B5-007} & Attach a PDF containing a highlight annotation. & Record it as annotated. \\[2pt]
\mbox{B5-008} & Attach a PDF containing a link annotation only. & Record it as unannotated. \\[2pt]
\mbox{B5-009} & Override a clean PDF as annotated; refresh fingerprints. & Preserve the manual annotated state. \\[2pt]
\mbox{B5-010} & Override a text-annotated PDF as unannotated; reopen and refresh. & Preserve the override after reopening. \\[2pt]
\midrule
\casefamilyfirst{B6\quad Preservation and duplicate reuse (14 cases)}
  {B6-001}
  {Attach identical PDF bytes twice to one paper.}
  {Reuse the resource ID; retain one resource.}
\mbox{B6-002} & Register distinct preprint and accepted-manuscript files. & Retain both classifications and original file bytes. \\[2pt]
\mbox{B6-003} & Register distinct accepted-manuscript and published files. & Retain both classifications and original file bytes. \\[2pt]
\mbox{B6-004} & Register a published version, corrected version and correction notice. & Retain all three classifications and file contents. \\[2pt]
\mbox{B6-005} & Register a published article, supplement and peer-review file. & Retain all three classifications and file contents. \\[2pt]
\mbox{B6-006} & Refresh metadata with a PDF URL but no downloaded file. & Create no file resource. \\[2pt]
\mbox{B6-007} & Attach equal-sized PDFs containing different bytes. & Retain two resources. \\[2pt]
\mbox{B6-008} & Register the same stored path with a redundant path component. & Reuse the existing resource ID. \\[2pt]
\mbox{B6-009} & Reattach identical bytes under a different filename. & Retain one resource and preserve the original file. \\[2pt]
\mbox{B6-010} & Attach identical bytes to two different paper objects. & Keep a resource associated with each paper. \\[2pt]
\mbox{B6-011} & Attach different bytes under the same filename. & Retain two resources. \\[2pt]
\mbox{B6-012} & Reattach identical bytes while a manual Default is active. & Retain one resource and the manual Default. \\[2pt]
\mbox{B6-013} & Reopen the database before attaching identical bytes again. & Reuse the original resource ID. \\[2pt]
\mbox{B6-014} & Attach identical bytes concurrently through two connections. & Retain one resource. \\[2pt]
\midrule
\casefamilyfirst{B7\quad File validation (14 cases)}
  {B7-001}
  {Validate an empty input.}
  {Reject it as an invalid PDF.}
\mbox{B7-002} & Validate plain text supplied as a PDF. & Reject it as an invalid PDF. \\[2pt]
\mbox{B7-003} & Validate a PDF header without document content. & Reject it as an invalid PDF. \\[2pt]
\mbox{B7-004} & Validate a header, empty object and end marker without pages. & Reject it as an invalid PDF. \\[2pt]
\mbox{B7-005} & Validate a PDF truncated at half its length. & Reject it as an invalid PDF. \\[2pt]
\mbox{B7-006} & Validate a PDF with its end-of-file marker removed. & Reject it as an invalid PDF. \\[2pt]
\mbox{B7-007} & Validate a prefixed PDF fragment with an added end marker. & Reject it as an invalid PDF. \\[2pt]
\mbox{B7-008} & Validate a complete one-page PDF. & Accept it as a valid PDF. \\[2pt]
\mbox{B7-009} & Validate a complete two-page PDF. & Accept it as a valid PDF. \\[2pt]
\mbox{B7-010} & Validate an encrypted PDF readable with an empty password. & Accept it as a valid PDF. \\[2pt]
\mbox{B7-011} & Validate a PDF locked with a non-empty password. & Reject it as unreadable. \\[2pt]
\mbox{B7-012} & Validate an HTML login page supplied as a PDF. & Reject it as an invalid PDF. \\[2pt]
\mbox{B7-013} & Validate a PDF containing zero pages. & Reject it as an invalid PDF. \\[2pt]
\mbox{B7-014} & Validate a PDF with an invalid cross-reference table and root. & Reject it as an invalid PDF. \\[2pt]
\midrule
\casefamilyfirst{B8\quad Source provenance (14 cases)}
  {B8-001}
  {Classify a download URL on arxiv.org.}
  {Record the source as a preprint server.}
\mbox{B8-002} & Classify a download URL on biorxiv.org. & Record the source as a preprint server. \\[2pt]
\mbox{B8-003} & Classify a download URL on repository.example.edu. & Record the source as an institutional repository. \\[2pt]
\mbox{B8-004} & Classify a download URL on eprints.example.ac.uk. & Record the source as an institutional repository. \\[2pt]
\mbox{B8-005} & Classify a nature.com URL acquired by publisher download. & Record the source as publisher. \\[2pt]
\mbox{B8-006} & Classify a science.org URL acquired by the AAAS download route. & Record the source as publisher. \\[2pt]
\mbox{B8-007} & Classify a manual import without a source URL. & Record the source as user import. \\[2pt]
\mbox{B8-008} & Classify an inventory import without a source URL. & Record the source as user import. \\[2pt]
\mbox{B8-009} & Classify a download from files.example.net. & Record the source as other. \\[2pt]
\mbox{B8-010} & Classify an input with no source URL or acquisition route. & Record the source as unknown. \\[2pt]
\mbox{B8-011} & Classify a browser capture without a source URL. & Record the source as user import. \\[2pt]
\mbox{B8-012} & Classify a URL on arxiv.org.evil.example. & Record other, not preprint server. \\[2pt]
\mbox{B8-013} & Classify a URL on nature.com.evil.example. & Record other, not publisher. \\[2pt]
\mbox{B8-014} & Classify a download URL on medrxiv.org. & Record the source as a preprint server. \\[2pt]
\casefamilyfirst{B9\quad Metadata and identifier correction (12 cases)}
  {B9-001}
  {Replace the DOI explicitly.}
  {Save the new DOI; preserve the paper, file and manual Default.}
\mbox{B9-002} & Clear the DOI explicitly. & Clear the identifier; preserve the paper, file and manual Default. \\[2pt]
\mbox{B9-003} & Replace the DOI with one owned by another paper. & Reject the collision and leave the original record unchanged. \\[2pt]
\mbox{B9-004} & Submit a malformed DOI. & Reject it and leave the original record unchanged. \\[2pt]
\mbox{B9-005} & Save a DOI correction before source confirmation. & Mark it pending; preserve the paper, file and manual Default. \\[2pt]
\mbox{B9-006} & Refresh a corrected DOI with matching supplied source metadata. & Update the bibliography; preserve the paper, file and manual Default. \\[2pt]
\mbox{B9-007} & Return source metadata carrying a different DOI. & Keep the earlier bibliography, file and manual Default. \\[2pt]
\mbox{B9-008} & Return no source metadata for the corrected DOI. & Keep the earlier bibliography, file and manual Default. \\[2pt]
\mbox{B9-009} & Repeat the same matching metadata refresh. & Keep one correction-history entry and preserve the file and Default. \\[2pt]
\mbox{B9-010} & Change the DOI manually while an earlier refresh is pending. & Discard the stale update; retain the newer DOI and existing file. \\[2pt]
\mbox{B9-011} & Add an arXiv identifier explicitly. & Save the identifier; preserve the paper, file and manual Default. \\[2pt]
\mbox{B9-012} & Run Quick refresh after an explicit DOI correction. & Apply matching metadata through the same correction service. \\[2pt]
\midrule
\casefamilyfirst{B10\quad MCP import and authorization (12 cases)}
  {B10-001}
  {Import supplied metadata through an authorized MCP connection.}
  {Create a paper record without a placeholder file.}
\mbox{B10-002} & Repeat an identical metadata import request. & Reuse the same Gewu Key and placement; create no file. \\[2pt]
\mbox{B10-003} & Append a PDF after a metadata-only import. & Retain the Gewu Key and register one file. \\[2pt]
\mbox{B10-004} & Import with a missing or invalid upload reference. & Reject the import; do not silently create a metadata-only record. \\[2pt]
\mbox{B10-005} & Import into a destination outside the write grant. & Reject the request without creating a paper. \\[2pt]
\mbox{B10-006} & Attempt to adopt an existing record in an unreadable private collection. & Reject the request without creating a paper. \\[2pt]
\mbox{B10-007} & Import a matching, uncollected provider record. & Reuse its existing Gewu Key rather than create another paper. \\[2pt]
\mbox{B10-008} & Import a new publication file while a manual Default is active. & Retain both files, the Gewu Key and the manual Default. \\[2pt]
\mbox{B10-009} & Import metadata missing required author information. & Reject it rather than perform an implicit lookup. \\[2pt]
\multicolumn{3}{@{}p{\linewidth}@{}}{%
\begin{minipage}[t]{\linewidth}
\begin{tabular}[t]{@{}>{\raggedright\arraybackslash}p{0.12\linewidth}
                       >{\raggedright\arraybackslash}p{0.43\linewidth}
                       >{\raggedright\arraybackslash}p{\dimexpr0.45\linewidth-16pt\relax}@{}}
\mbox{B10-010} & Reuse a request ID with changed metadata. & Report a conflict and preserve the original metadata. \\[2pt]
\mbox{B10-011} & Import through a revoked connection. & Reject the request without creating a paper. \\[2pt]
\mbox{B10-012} & Import through a papers-read-only connection. & Reject the write without creating a paper. \\[2pt]
\caseboundaryrule
\end{tabular}\par\vspace{4pt}
{\normalfont\rmfamily\GewuCaptionFont
\textbf{B1 decision rules (revision 539d00d).}
DOI strings are normalized; differing explicit DOIs and DOI/URL inconsistencies
prevent automatic merging. A title conflict requires both title-token Jaccard
similarity $<0.22$ and Python's {SequenceMatcher} ratio $<0.48$ after
case normalization and alphanumeric tokenization. It blocks acceptance only
when accompanied by disjoint author-name keys or incompatible journal names.
Author keys use the final token of each normalized name. Non-placeholder
journal names conflict when neither contains the other and their token sets
are disjoint. Missing values do not themselves establish a conflict.
In B1-017, title agreement without an identifier match leaves the
storage records separate. Author Monitor's separate fallback permits an exact-title
match (ignoring case) when identifier matching fails, conflict checks pass and
publication years agree when both are present; missing years do not block reuse.
This heuristic does not independently confirm authorship and is not exercised by B1-017.\par}
\end{minipage}
}\\
\end{longtable}

\clearpage
\renewcommand{\small}{\fontsize{10}{12}\selectfont}
\small
\setlength{\tabcolsep}{4pt}
\setlength{\aboverulesep}{2pt}
\setlength{\belowrulesep}{2pt}
\renewcommand{\arraystretch}{1.04}
\begin{longtable}{@{}>{\raggedright\arraybackslash}p{0.12\linewidth}
                       >{\raggedright\arraybackslash}p{0.43\linewidth}
                       >{\raggedright\arraybackslash}p{\dimexpr0.45\linewidth-16pt\relax}@{}}
\caption{\textbf{Individual state-continuity scenarios (Task C).}
All 104 scenarios passed. Each family contains four distinct scenarios
(V01-V04), not repeated runs. A pass requires every applicable operation and
check to succeed. The rows summarize the operations and expected states; full
definitions and recorded outputs accompany the evaluation data.}
\label{tab:task-c-scenarios}\\
\caseboundaryrule
\textbf{ID} & \textbf{Input or operation} & \textbf{Expected outcome} \\
\caseboundaryrule
\endfirsthead
\multicolumn{3}{@{}l}{Supplementary Table~\thetable\ (continued)}\\
\caseboundaryrule
\textbf{ID} & \textbf{Input or operation} & \textbf{Expected outcome} \\
\caseboundaryrule
\endhead
\caseboundaryrule
\multicolumn{3}{r@{}}{Continued on next page}\\
\endfoot
\caseboundaryrule
\endlastfoot
\casefamilyfirst{C01\quad Multi-channel identity (4 scenarios)}
  {C01-V01}
  {Ingest the same DOI through Journal, Subject and Author Monitor.}
  {Retain one paper with all three discovery associations.}
\mbox{C01-V02} & Ingest an arXiv revision, then DOI-bearing journal and author records. & Retain one paper with the accumulated discovery associations. \\[2pt]
\mbox{C01-V03} & Ingest a title-only Scholar record, then journal and subject identifiers. & Reconcile to one paper with all three discovery associations. \\[2pt]
\mbox{C01-V04} & Ingest a DOI-bearing journal record, then Scholar and subject records. & Retain one paper with all three discovery associations. \\[2pt]
\midrule
\casefamilyfirst{C02\quad Preprint to publication (4 scenarios)}
  {C02-V01}
  {Add a published version after a preprint.}
  {Retain both files and their version relation; select the published file.}
\mbox{C02-V02} & Add a later preprint revision, then a published version. & Retain the earlier files and link the published version to a preprint. \\[2pt]
\mbox{C02-V03} & Refresh publisher metadata between preprint and published-file acquisition. & Retain both versions and valid paths; select the published file. \\[2pt]
\mbox{C02-V04} & Acquire the published version before the preprint. & Retain both files; the late preprint does not replace the Default. \\[2pt]
\midrule
\casefamilyfirst{C03\quad Accepted to published version (4 scenarios)}
  {C03-V01}
  {Add a published version after an accepted manuscript.}
  {Retain both versions and their relation; select the published file.}
\mbox{C03-V02} & Acquire a preprint, accepted manuscript and published version in sequence. & Retain the accepted and published files with the published file selected. \\[2pt]
\mbox{C03-V03} & Acquire the same accepted manuscript twice, then the published version. & Reuse the duplicate accepted file and retain the published file separately. \\[2pt]
\mbox{C03-V04} & Acquire the published version before the accepted manuscript. & Retain both versions; keep the published file selected. \\[2pt]
\midrule
\casefamilyfirst{C04\quad Repeated acquisition (4 scenarios)}
  {C04-V01}
  {Reattach identical bytes repeatedly under the same filename.}
  {Reuse one resource and its selection; leave no orphan managed file.}
\mbox{C04-V02} & Reattach identical bytes under different filenames. & Reuse one resource and its selection; leave no orphan managed file. \\[2pt]
\mbox{C04-V03} & Repeat forced downloads containing identical bytes. & Reuse one resource and its selection; leave no orphan managed file. \\[2pt]
\mbox{C04-V04} & Supply identical publisher copies for a file initially marked as preprint. & Reuse the existing resource across both publisher acquisitions. \\[2pt]
\midrule
\casefamilyfirst{C05\quad Clean and annotated copies (4 scenarios)}
  {C05-V01}
  {Attach a clean PDF followed by its annotated copy.}
  {Retain both files with their content relation and distinct annotation states.}
\mbox{C05-V02} & Attach the annotated copy before the clean PDF. & Retain both files with their content relation and distinct annotation states. \\[2pt]
\mbox{C05-V03} & Attach both copies and explicitly refresh their fingerprints. & Preserve the two resources and their content relation after refresh. \\[2pt]
\mbox{C05-V04} & Attach the annotated copy first, then explicitly correct its annotation role. & Preserve both resources and the corrected annotation state. \\[2pt]
\midrule
\casefamilyfirst{C06\quad Secondary-file selection (4 scenarios)}
  {C06-V01}
  {Add a publication, supplement and note; select the supplement, then clear.}
  {Honor the manual selection; restore the publication and retain all files.}
\mbox{C06-V02} & Add a publication, other PDF and note; select the other PDF, then clear. & Honor the manual selection; restore the publication and retain all files. \\[2pt]
\mbox{C06-V03} & Add a publication, supplement and note; select the note, then clear. & Honor the manual selection; restore the publication and retain all files. \\[2pt]
\mbox{C06-V04} & Add a publication, supplement and note without a manual override. & Select the publication automatically and retain all three resources. \\[2pt]
\midrule
\casefamilyfirst{C07\quad Manual and automatic Default (4 scenarios)}
  {C07-V01}
  {Manually select the preprint from a version set, then clear the choice.}
  {Honor the choice and restore automatic publication selection after clearing.}
\mbox{C07-V02} & Manually select the accepted manuscript, then clear the choice. & Honor the choice and restore automatic publication selection after clearing. \\[2pt]
\mbox{C07-V03} & Manually select the current published version, then clear the choice. & Preserve the selected file and correctly clear the manual-selection flag. \\[2pt]
\mbox{C07-V04} & Add a new published version while a manual choice is active, then clear. & Keep the manual choice until cleared; then apply automatic ranking. \\[2pt]
\midrule
\casefamilyfirst{C08\quad Multiple placements (4 scenarios)}
  {C08-V01}
  {Place one paper in two Library folders and synchronize its file.}
  {Retain one identity, both placements and valid paths after reopening.}
\mbox{C08-V02} & Place one paper in Library and a Space; synchronize its file. & Retain one identity, both placements and valid paths after reopening. \\[2pt]
\mbox{C08-V03} & Place one paper in Library and My Publications; synchronize its file. & Retain one identity, both associations and valid paths after reopening. \\[2pt]
\mbox{C08-V04} & Place one paper in two Spaces and My Publications; synchronize its file. & Retain one identity, all associations and valid paths after reopening. \\[2pt]
\midrule
\casefamilyfirst{C09\quad Scoped removal (4 scenarios)}
  {C09-V01}
  {Remove one Library placement from a paper shared across contexts.}
  {Preserve the other associations and retained file resources.}
\mbox{C09-V02} & Remove an Author Monitor association from a shared paper. & Preserve the other associations and retained file resources. \\[2pt]
\mbox{C09-V03} & Remove a My Publications association from a shared paper. & Preserve the other associations and retained file resources. \\[2pt]
\mbox{C09-V04} & Remove a non-Default file relation from a shared paper. & Preserve unrelated associations and the remaining resources. \\[2pt]
\midrule
\casefamilyfirst{C10\quad Paths and restart (4 scenarios)}
  {C10-V01}
  {Move a managed paper placement to another folder.}
  {Synchronize paths; retain resources without orphan records or files.}
\mbox{C10-V02} & Rename a managed paper item. & Synchronize paths; retain resources without orphan records or files. \\[2pt]
\mbox{C10-V03} & Move a managed paper item, then rename it. & Preserve the resource set and valid paths through both operations. \\[2pt]
\mbox{C10-V04} & Move and rename an item, then repair an external filesystem rename. & Restore valid paths and retain the resource set after reopening. \\[2pt]
\midrule
\casefamilyfirst{C11\quad Failure and recovery (4 scenarios)}
  {C11-V01}
  {Submit HTML as a PDF, then retry with a valid PDF.}
  {Reject the invalid input; retain prior state and complete the valid retry.}
\mbox{C11-V02} & Submit an empty PDF, then retry with a valid PDF. & Reject the invalid input; retain prior state and complete the valid retry. \\[2pt]
\mbox{C11-V03} & Submit a truncated PDF header, then retry with a valid PDF. & Reject the invalid input; retain prior state and complete the valid retry. \\[2pt]
\mbox{C11-V04} & Submit a file with a late PDF marker, then retry with a valid PDF. & Reject the invalid input; retain prior state and complete the valid retry. \\[2pt]
\midrule
\casefamilyfirst{C12\quad Schema migration (4 scenarios)}
  {C12-V01}
  {Open a schema-5 database requiring primary-resource backfill.}
  {Migrate to the current schema; preserve the paper, file and folder mapping.}
\mbox{C12-V02} & Open a schema-6 database with nested folders and legacy file state. & Migrate while preserving file bytes, resource count and valid paths. \\[2pt]
\mbox{C12-V03} & Open a schema-7 database with an existing primary resource. & Migrate while retaining that resource, file bytes and valid paths. \\[2pt]
\mbox{C12-V04} & Open a schema-8 database containing a primary file and supplement. & Migrate while retaining both resources, their bytes and valid paths. \\[2pt]
\midrule
\casefamilyfirst{C13\quad Workspace merge and rollback (4 scenarios)}
  {C13-V01}
  {Merge a standalone object containing one file into another paper.}
  {Transfer the file and references consistently; retain unrelated state.}
\mbox{C13-V02} & Merge a standalone object containing several files into another paper. & Transfer all files and references consistently; retain unrelated state. \\[2pt]
\mbox{C13-V03} & Attempt a merge whose source is shared across placements. & Block the unsafe operation and preserve the prior state. \\[2pt]
\mbox{C13-V04} & Inject a failure during the second file move of a merge. & Roll back both record and file changes without losing resources. \\[2pt]
\midrule
\casefamilyfirst{C14\quad Publication-notice processing (4 scenarios)}
  {C14-V01}
  {Process a journal feed with a notice first, then repeat the feed.}
  {Retain two research papers and one notice event without duplication.}
\mbox{C14-V02} & Process a journal feed with a notice in the middle, then repeat it. & Retain two research papers and one notice event without duplication. \\[2pt]
\mbox{C14-V03} & Process a journal feed with a notice last, then repeat the feed. & Retain two research papers and one notice event without duplication. \\[2pt]
\mbox{C14-V04} & Process and repeat a feed containing five notices only. & Retain five notice events, no ordinary papers and no duplicate identities. \\[2pt]
\midrule
\casefamilyfirst{C15\quad Notice import and association (4 scenarios)}
  {C15-V01}
  {Import a browser-acquired notice for one explicitly identified paper.}
  {Require the target; retain the notice separately and preserve the Default.}
\mbox{C15-V02} & Load a notice file without an identified target. & Retain the file awaiting association without creating an ordinary paper. \\[2pt]
\mbox{C15-V03} & Import a notice through MCP for one explicit target and retry. & Reuse the same paper and notice file; preserve the original and Default. \\[2pt]
\mbox{C15-V04} & Attempt unknown, multiple and self-targets, then supply a valid target. & Reject invalid relations without attaching files; accept the corrected retry. \\[2pt]
\midrule
\casefamilyfirst{C16\quad Source failure and cache (4 scenarios)}
  {C16-V01}
  {Receive some successful source responses and one failure; use cache, then retry.}
  {Keep successful entries, report partial status and complete the retry.}
\mbox{C16-V02} & Receive an unchanged-source response after a complete successful retrieval. & Reuse the complete cache and retain all three entries. \\[2pt]
\mbox{C16-V03} & Retry a source with a partial or failed cached result. & Request the missing response body and preserve previously retrieved entries. \\[2pt]
\mbox{C16-V04} & Let one source fail before querying another source. & Continue to the next source and retain its results with partial status. \\[2pt]
\midrule
\casefamilyfirst{C17\quad Canonical identity and aliases (4 scenarios)}
  {C17-V01}
  {Ingest one DOI through different entry points, then read through MCP.}
  {Retain one canonical identity accessible through its Gewu Key.}
\mbox{C17-V02} & Write one DOI concurrently, then read through MCP. & Retain one canonical identity accessible through its Gewu Key. \\[2pt]
\mbox{C17-V03} & Establish a DOI/arXiv relationship after initial ingestion. & Reconcile identity and preserve access through the canonical key. \\[2pt]
\mbox{C17-V04} & Read a reconciled paper using an earlier key, then test a restricted grant. & Resolve the earlier key without expanding access beyond authorization. \\[2pt]
\midrule
\casefamilyfirst{C18\quad Identifier and metadata updates (4 scenarios)}
  {C18-V01}
  {Replace a DOI explicitly, refresh metadata and read through MCP.}
  {Apply matching metadata; preserve the key, file and manual Default.}
\mbox{C18-V02} & Clear and refill the DOI; receive a mismatching source response. & Keep the earlier bibliography and retain access through the same key. \\[2pt]
\mbox{C18-V03} & Change the identifier while an earlier metadata response is pending. & Discard the stale update; preserve the newer identifier and existing file. \\[2pt]
\mbox{C18-V04} & Add an arXiv identifier, then attempt a conflicting DOI replacement. & Save the valid identifier; reject the collision without changing the record. \\[2pt]
\midrule
\casefamilyfirst{C19\quad Metadata task control (4 scenarios)}
  {C19-V01}
  {Pause a metadata task during processing, then resume it.}
  {Continue from its checkpoint; preserve the key, file and manual Default.}
\mbox{C19-V02} & Restart after a metadata task pauses, then resume it. & Retain progress across restart and complete the remaining work. \\[2pt]
\mbox{C19-V03} & Remove an item from the authorized folder while the task is paused. & Recheck membership on resumption and preserve existing file state. \\[2pt]
\mbox{C19-V04} & Combine Quick and Full runs with a partial-source retry. & Retry the failed source, avoid repeating successful work and retain files. \\[2pt]
\midrule
\casefamilyfirst{C20\quad Browser file acquisition (4 scenarios)}
  {C20-V01}
  {Review a browser-acquired PDF for an existing work, then save it twice.}
  {Reuse the key and duplicate file; preserve the manual Default.}
\mbox{C20-V02} & Review a browser-acquired PDF, then cancel the preview. & Leave records and resources unchanged and remove the temporary capture. \\[2pt]
\mbox{C20-V03} & Change the identifier manually after preview but before saving. & Reject the stale save without registering a file. \\[2pt]
\mbox{C20-V04} & Replace the previewed PDF with one carrying conflicting identifiers. & Require renewed review and prevent the conflicting file from being saved. \\[2pt]
\midrule
\casefamilyfirst{C21\quad Metadata-only to file import (4 scenarios)}
  {C21-V01}
  {Import metadata only, then append a PDF and retry.}
  {Create no placeholder; retain one key and one file after retry.}
\mbox{C21-V02} & Import a paper already known to a Monitor, append a PDF, then retry the file attachment. & Reuse the Monitor's Gewu Key and retain one file after retry. \\[2pt]
\mbox{C21-V03} & Repeat an import request, then reuse its ID with changed parameters. & Reuse an identical request; reject changed parameters without duplication. \\[2pt]
\mbox{C21-V04} & Submit an invalid upload, then valid metadata and a PDF. & Reject the invalid upload without fallback; accept the later valid import. \\[2pt]
\midrule
\casefamilyfirst{C22\quad AI authorization changes (4 scenarios)}
  {C22-V01}
  {Read through a read-only connection, then attempt a write.}
  {Permit the read, reject the write and preserve identity.}
\mbox{C22-V02} & Read successfully, then revoke or expire the connection. & Reject subsequent paper-metadata requests through the revoked or expired connection. \\[2pt]
\mbox{C22-V03} & Read a paper, remove its authorized folder membership and read again. & Reject the later read after rechecking membership. \\[2pt]
\mbox{C22-V04} & Attempt to reach an ungranted paper through aliases or a shared file path. & Do not expand the connection's readable scope. \\[2pt]
\midrule
\casefamilyfirst{C23\quad File delivery and analysis append (4 scenarios)}
  {C23-V01}
  {Deliver a file, append an analysis and read it from a new session.}
  {Deliver a read-only copy; retain the analysis, original and existing Default.}
\mbox{C23-V02} & Repeat an analysis-save request, then change its content under the same ID. & Reuse the first analysis; reject conflicting content without overwriting it. \\[2pt]
\mbox{C23-V03} & Change the source bytes after delivery and before saving an analysis. & Reject the stale analysis and preserve the previously delivered copy. \\[2pt]
\mbox{C23-V04} & Exhaust and release delivery quota; append another analysis and attempt overwrite. & Restore delivery capacity; preserve original files and earlier analyses. \\[2pt]
\midrule
\casefamilyfirst{C24\quad MCP file organization (4 scenarios)}
  {C24-V01}
  {Rename a managed file through MCP.}
  {Update its name while retaining the resource ID and file bytes.}
\mbox{C24-V02} & Move an authorized directory containing managed and unregistered content. & Preserve resource identity, empty directories and other user files. \\[2pt]
\mbox{C24-V03} & Attempt to move content across an unauthorized or shared boundary. & Reject the move; preserve membership and file bytes. \\[2pt]
\mbox{C24-V04} & Inject a failure during a directory move, then retry. & Roll back record changes and file moves; complete the valid retry. \\[2pt]
\midrule
\casefamilyfirst{C25\quad Physical file ownership (4 scenarios)}
  {C25-V01}
  {Change which workspace root owns files shared across placements.}
  {Retain one physical copy per resource, valid links and the same paper key.}
\mbox{C25-V02} & Check a Finder alias and rename its target in the filesystem. & Retain native alias behavior, a resolvable target and unchanged file bytes. \\[2pt]
\mbox{C25-V03} & Transfer a file from an externally linked standalone record. & Preserve the external original and retain a managed copy under its new owner. \\[2pt]
\mbox{C25-V04} & Inject Trash and file-operation failures, recover after restart and retry. & Restore registered bytes and records; complete the retry without changing Default. \\[2pt]
\midrule
\casefamilyfirst{C26\quad My Publications state (4 scenarios)}
  {C26-V01}
  {Pause a paginated My Publications import and resume it.}
  {Resume at the saved position and retain three unique placements.}
\mbox{C26-V02} & Restart after a paginated import pauses, then resume it. & Retain the saved position and complete without duplicate placements. \\[2pt]
\mbox{C26-V03} & Delete an imported placement before continuing the import. & Respect the deletion marker; do not recreate that placement. \\[2pt]
\mbox{C26-V04} & Encounter a source restriction, then resume the import. & Report the restriction without advancing progress; complete the later retry. \\[2pt]
\end{longtable}

\clearpage
\begingroup
\fontsize{10}{12}\selectfont
\setlength{\tabcolsep}{4pt}
\renewcommand{\arraystretch}{1.04}
\refstepcounter{table}\label{tab:ibsc-interpretation-trace}
{\normalfont\rmfamily\GewuCaptionFont\noindent
\textbf{Supplementary Table~\thetable. Traceable refinement of an interpretation in the IBSC case study.}\par}
\vspace{5pt}
\noindent The retained Cheriton et al. record\cite{Cheriton2020TwoPhoton}
connects an earlier analysis, source rereading, an appended follow-up and
reuse in an Obsidian synthesis.\par\vspace{6pt}
\noindent\begin{tabular}{@{}>{\raggedright\arraybackslash}p{0.12\linewidth}
                            >{\raggedright\arraybackslash}p{0.59\linewidth}
                            >{\raggedright\arraybackslash}p{\dimexpr0.29\linewidth-16pt\relax}@{}}
\toprule
\textbf{Stage} & \textbf{Recorded interpretation or action} & \textbf{Evidence and record locator} \\
\midrule
1. Earlier analysis &
The saved note already attributed the no-bias subgap response mainly to leakage,
recognized unexplained room-temperature above-gap enhancement, and distinguished
44\% photocurrent enhancement from 0.016\% efficiency at 4 suns. Its overview
nevertheless used broad exclusion language for heating, thermal escape and tunneling. &
R2: earlier analysis, file 1124; Overview, Evidence chain (Fig.~3d), Main advance and Limitations. \\
\addlinespace[6pt]
2. Return to the source &
Fig.~3d compares external quantum efficiency (EQE) with and without 850-nm bias
at 78 and 295 K. Fig.~3e is a temperature scan without light bias, providing
indirect evidence against a heating-dominated explanation. Supplementary
Fig.~6a uses weak 635-nm bias; Fig.~6b gives no-bias temperature spectra from
a different growth. &
R1: main pp.~3-5, Fig.~3d,e (p.~5); Supplementary Fig.~6a,b (supplement p.~3;
merged PDF p.~10). R3, section~2, records the reread. \\
\addlinespace[6pt]
3. Refined interpretation &
The follow-up makes the baseline/increment distinction explicit and qualifies
the thermal controls as indirect evidence, not direct laser thermometry or a
unique decomposition of room-temperature current. Combined evidence supports
a two-step contribution. The old pairing of 58$\to$114 $\mu$A/cm$^2$ with
44\% was corrected, while the paper's prose/plot discrepancy remained
unresolved. The photocurrent/efficiency distinction was retained. &
R2 versus R3: follow-up file 1410; Central assessment, sections~1-4, and
Additions and corrections. Fig.~3f estimates are not raw data. \\
\addlinespace[6pt]
4. Append and retain &
The follow-up was appended on 27 September 2026 as file 1410 under the same
paper. Earlier file 1124 remained separately retrievable. Delivered-file hashes
match the historical source references and save receipt; the follow-up did not
replace the earlier analysis. &
Gewu Key: gw\_f81b2ade5a. R4: save receipt and file 1410 provenance.
R2/R3: earlier and appended analyses. \\
\addlinespace[6pt]
5. Reuse in synthesis &
The Obsidian synthesis retains the no-bias baseline and combined-evidence
interpretation while separating mechanism attribution from device efficiency.
It explicitly cites file 1410 and asks how selective two-beam response connects
to J-V/maximum-power changes on the same device, with auxiliary-light input counted. &
R5: synthesis note; Cross-paper comparison, Specific revisions and Next research
questions. Citations include DOI, Gewu Key, figure/pages and file ID 1410. \\
\bottomrule
\end{tabular}
\par\vspace{6pt}
\noindent R1, original article and supplement (file 1102); R2, earlier analysis
(file 1124); R3, appended follow-up (file 1410); R4, save record; R5,
\textit{Distinguishing evidence for two-step absorption and device efficiency}.
\par
\endgroup

\endgroup

\end{document}